\documentclass[aps,nofootinbib,twocolumn,floatfix,prd,superscriptaddress]{revtex4} %eqsecnum,
\usepackage{graphicx,amsmath,amssymb,color,bm,hyperref,epsfig}
\def\be{\begin{equation}}
\def\ee{\end{equation}}
\def\bea{\begin{eqnarray}}
\def\eea{\end{eqnarray}}
\newcommand{\vs}{\nonumber\\}
\def\ba#1\ea{\begin{align}#1\end{align}}

\newcommand{\fsky}{f_{\rm sky}}

\newcommand{\g}{\gamma}
\newcommand{\refeq}[1]{Eq.~(\ref{eq:#1})}          
\newcommand{\refeqs}[2]{Eqs.~(\ref{eq:#1})--(\ref{eq:#2})}          
          
\newcommand{\reffig}[1]{Fig.~\ref{fig:#1}}          

\newcommand{\refsec}[1]{\S~\ref{sec:#1}}          
\newcommand{\refapp}[1]{App.~\ref{app:#1}}

\renewcommand{\v}[1]{\mathbf{#1}}
\newcommand{\vx}{\v{x}}
\newcommand{\vk}{\v{k}}

\newcommand{\<}{\langle}
\renewcommand{\>}{\rangle}

\renewcommand{\k}{\kappa}
\renewcommand{\d}{\delta}
\newcommand{\D}{\Delta}

\newcommand{\nhat}{\hat{n}}
\newcommand{\vnhat}{\v{\hat{n}}}

\newcommand{\eps}{\varepsilon}

\renewcommand{\a}{\alpha}
\renewcommand{\b}{\beta}

\newcommand{\rhocr}{\rho_{\rm cr 0}}
\newcommand{\zt}{\tilde{z}}

\newcommand{\chit}{\tilde{\chi}}
\newcommand{\Mpch}{\:{\rm Mpc}/h}
\newcommand{\iMpch}{\:h/{\rm Mpc}}

\def\A{\mathcal{A}}
\def\P{\mathcal{P}}

\def\O{\mathcal{O}}
\def\Del{\eth}
\renewcommand{\Re}{{\rm Re}\,}
\renewcommand{\Im}{{\rm Im}\,}

\def\G{\Gamma}

\def\nhat{\hat{n}}
\def\vnhat{\hat{\v{n}}}

\begin{document}

\title{Large-Scale Structure with Gravitational Waves II: Shear}

\author{Fabian Schmidt}
\affiliation{Theoretical Astrophysics,
	California Institute of Technology, 
	Mail Code 350-17, Pasadena, CA  91125, USA}
\author{Donghui Jeong}
\affiliation{Department of Physics and Astronomy, Johns Hopkins University, 3400 N. Charles St., Baltimore, MD 21210, USA}

\begin{abstract}
The $B$- (curl-)mode of the correlation of galaxy ellipticities (shear) 
can be used to detect a stochastic gravitational wave background, such 
as that predicted by inflation.  
In this paper, we derive the tensor mode contributions to shear 
from both gravitational lensing and intrinsic alignments, using the 
gauge-invariant, full-sky results of \citet{stdruler}.  
We find that the intrinsic alignment contribution, calculated using the linear
alignment model, is larger than the lensing contribution
by an order of magnitude or more, if the alignment strength for tensor
modes is of the same order as for scalar modes.  This contribution also
extends to higher multipoles.  These results make the prospects for
probing tensor modes using galaxy surveys less pessimistic than previously
thought, though still very challenging.  
\end{abstract}

\date{\today}

\pacs{98.65.Dx, 98.65.-r, 98.80.Jk}

\maketitle

%%%%%%%%%%%%%%%%%%%%%%%%%%%%%%%%%%%%%%%%%%%%%%%%%%%%%%%%%%%%%%%%%%%%%%%%%%%
%%%%%%%%%%%%%%%%%%%%%%%%%%%%%%%%%%%%%%%%%%%%%%%%%%%%%%%%%%%%%%%%%%%%%%%%%%%
\section{Introduction}
\label{sec:intro}

A stochastic gravitational wave (GW) background
is one of the key testable predictions of inflation.  Thus, much
theoretical and experimental effort is devoted to searching for this
gravitational wave background.  
One necessary ingredient of any method designed to search for
a GW background is the ability to cleanly separate the GW contribution
from scalar perturbations.  Most commonly, this is done by considering
spin-2 quantities on the sky, such as 
the anisotropy in the polarization of the cosmic microwave background (CMB)
radiation \cite{SelZalPRL,KamionkowskiKosowskyStebbins},
the anisotropy of the 2-point correlation function 
\cite{Cooray04,MasuiPen,BookEtal,paperI}, or the ellipticity 
of galaxy images \cite{DodelsonEtal}.  
The polarization of the cosmic microwave background is commonly 
considered as the most promising probe.  However, the 21cm emission
from the dark ages has recently been shown to in principle offer even more
discovery potential \cite{MasuiPen,BookMKFS}.  On the other hand,
previous authors have concluded that weak lensing shear will
most likely not be a competitive probe of primordial GW
\cite{DodelsonEtal,SarkarEtal08,Dodelson10}.  
Nevertheless, given the scientific impact, it will be crucial to confirm 
a possible detection of a GW background in the CMB with independent methods,
such as shear.  

The goal of this paper, and its companion \cite{paperI}, is to 
systematically and rigorously derive the GW effects on large-scale structure 
observables.  While we restrict ourselves to a linear treatment in the 
tensor perturbations, we strive to keep the results as general as possible
otherwise.  

This paper deals with shear, i.e. the correlations of galaxy ellipticities.  
The underlying assumption in interpreting shear correlations is that
in the absence of perturbations, galaxy ellipticities are uncorrelated
on large scales.  However, in the relativistic context, this raises
the question of the \emph{frame of reference} in which galaxy ellipticities
are actually uncorrelated.  In \citet{stdruler}, we have derived an
expression for the shear based strictly on observable quantities
(``standard rulers'').  As shown there, that expression is equivalent
to the statement that galaxies are isotropically oriented in their local
inertial frame, described by the Fermi normal
coordinates along a given galaxy's geodesic.  

Consider a region of spatial extent $R$, much larger than the size
of individual galaxies but smaller than the typical wavelength of the
perturbations we aim to measure through shear.  The center of mass of
this region moves on a time-like geodesic.  We can construct
a coordinate system where the center of mass is at rest at the origin,
and the time coordinate $t_{F}$ corresponds to the proper time of this
geodesic.  In other words, the unit vector defining the time coordinate is
equal to the tangent vector to this geodesic.  The spatial coordinate lines 
are chosen to be space-like geodesics (``straight lines'') orthogonal
to this time direction, and whose unit vectors are parallel-transported
along the observer's geodesic.  These are the so-called Fermi normal
coordinates (FNC) \cite{Fermi,ManasseMisner}.  
It is straightforward to construct them perturbatively for a given
metric and time-like geodesic (see \refapp{Fermi}).  
The most important property of Fermi normal coordinates is that the metric
is Minkowski at the spatial origin at all times $t_{F}$, with corrections
away from the geodesic going as $x_{F}^i x_{F}^j/R_c^2$, where $x_{F}^i$ denote the 
spatial Fermi coordinates and $R_c$ is the typical curvature scale of
the space-time ($R_c \sim H^{-1}$ for an unperturbed FRW space-time).  
Thus, the Fermi normal coordinates are the frame in which
a local observer in a weak gravitational field would describe her
experiments.  Neglecting the corrections $\propto x_{F}^2$, there
is no preferred direction in these coordinates along which galaxies
could align, and they thus have to be oriented isotropically.  
The contribution to the shear from the transformation to the local
Fermi coordinates was first introduced by \citet{DodelsonEtal}, who
showed that this term, which they call ``metric shear'', has a significant
impact on the observational signatures of tensor modes.  
Furthermore, as shown in \cite{stdruler}, this term is crucial to ensure
that the expression for the observable shear does not receive contributions from
constant and pure-gradient metric perturbations, as required by 
coordinate-invariance.  

Here, we evaluate this expression for the shear in an FRW space-time 
with tensor modes, which yields the
shear contribution by ``projection effects'' induced by a GW background.  
As discussed in \cite{KaiserJaffe,paperI}, the tensor contributions typically
peak at the source location rather than at lower redshifts, and there
is no enhancement of the contribution by transverse modes.  Both of
these facts lead to qualitative differences from the scalar case.

As discussed, the space-time in Fermi coordinates around a given galaxy 
is not perfectly flat however, and the corrections to the 
metric $\propto x_{F}^i x_{F}^j/R_c^2$ 
provide preferred directions along which galaxies can align.  
For non-relativistic motions (which generally applies to large-scale structure),
the relevant contribution to the metric in the Fermi frame corresponds to a 
tidal field, which can align galaxies and thus contribute to the observed
shear correlation.  
Here we derive the contribution to the tidal field by tensor
perturbations, and for the first time calculate the intrinsic alignment 
contribution of tensor modes in the linear alignment model.  This 
prescription has been shown to agree
well with observations on large scales in the scalar case.  
Note that this approach is only applicable for gravitational waves
with periods much longer than the dynamical time of galaxies.  

The key advantage of the shear applies to both lensing and linear intrinsic
alignment contributions:  linear scalar perturbations contribute only to 
the parity-even $E$-(gradient-)mode component at linear order, while tensor 
perturbations also 
contribute to the $B$-(curl-)mode (both through lensing and intrinsic 
alignments).  Importantly, scalar perturbations do contribute 
to $B$-modes at second and higher order, a point which we will discuss in
detail in \refsec{scalar}.

The outline of the paper is a follows: we introduce our notation
and conventions in \refsec{prelim} (they are the same as in the
companion paper \cite{paperI}).  
\refsec{lensing} presents the
derivation of the lensing (projection) contribution, while \refsec{IA} 
discusses the intrinsic alignment
contribution from tensor modes.  \refsec{Clshear} gives the expressions for the
$E-$ and $B$-mode power spectrum of the shear, including the connection
to previous results.  \refsec{res} presents
the results.  We conclude in \refsec{concl}.  
The appendix contains details on Fermi normal coordinates, the derivation
of angular power spectra, and the connection to convergence and rotation.  

%%%%%%%%%%%%%%%%%%%%%%%%%%%%%%%%%%%%%%%%%%%%%%%%%%%%%%%%%%%%%%%%%%%%%%%%%%%
%%%%%%%%%%%%%%%%%%%%%%%%%%%%%%%%%%%%%%%%%%%%%%%%%%%%%%%%%%%%%%%%%%%%%%%%%%%
\section{Preliminaries}
\label{sec:prelim}

We begin by introducing our convention for the metric and tensor perturbations
and some notation; it is identical to that used in \cite{paperI}.    
For simplicity, we restrict 
ourselves to a spatially flat FRW background, and consider only tensor
(gravitational wave) modes in the main part of the paper.  The perturbed 
metric is given by
\be
ds^2 
=
a^2(\eta)
\left[
-d\eta^2 + \left(\delta_{ij}+h_{ij} \right) dx^idx^j
\right],
\label{eq:metric}
\ee
where $h_{ij}$ is a metric perturbation which is assumed to be
transverse and traceless:
\be
h^i_{\; i} = 0 = (h_{ik})^{,i}.
\ee
We then decompose $h_{ij}$ into Fourier modes of two polarization states,
\ba
h_{ij}(\vk, \eta) =\:& e^+_{ij}(\hat\vk) h^+(\vk,\eta) + e^\times_{ij}(\hat\vk)
h^\times(\vk, \eta),
\label{eq:hpol}
\ea
where $e^s_{ij}(\hat\vk)$, $s=+,\times$, are transverse 
(with respect to $\hat\vk$) and 
traceless polarization tensors normalized through
$e^s_{ij} e^{s'\:ij} = 2 \d^{ss'}$.  Note that $h_s = e_s^{ij}h_{ij}/2$.
We assume both polarizations to
be independent and to have equal power spectra:
\ba
\< h_{s}(\vk,\eta) h_{s'}(\vk',\eta') \> =\:& (2\pi)^3 \d_D(\vk-\vk') 
\d_{ss'} \frac14 P_T(k,\eta,\eta').
\label{eq:PT}
\ea
Here, $\eta$ denotes conformal time, and the unequal-time power spectrum
is given by
\ba
P_{T}(k,\eta,\eta') =\:& T_T(k,\eta) T_T(k,\eta') P_{T0}(k),
\label{eq:PT2}
\ea
where $T_T(k,\eta)$ is the tensor transfer function, and the primordial
tensor power spectrum is specified through an amplitude $\Delta_T^2$ and
an index $n_T$ via
\ba
P_{T0}(k) =\:& 2\pi^2\,k^{-3} \left(\frac{k}{k_0}\right)^{n_T} \Delta_T^2.
\label{eq:PT0}
\ea
Following \emph{WMAP} convention \cite{komatsu/etal:2011},
we choose $k_0 = 0.002\:{\rm Mpc}^{-1}$ as pivot scale.  Throughout,
we will assume a scalar-to-tensor ratio of $r = 0.2$ at $k_0$ (consistent with
the 95\% confidence level WMAP bound), which together with our fiducial
cosmology determines $\Delta_T^2$.  The tensor index is chosen to follow
the inflationary consistency relation, $n_T = -r/8 = -0.0025$.  
For the expansion history,  we assume
a flat $\Lambda$CDM cosmology with $h=0.72$ and $\Omega_m=0.28$.  Contributions
from scalar perturbations are evaluated using a
spectral index of $n_s=0.958$ and power spectrum normalization at $z=0$ of
$\sigma_8 = 0.8$.

From \refeq{hpol} and \refeq{PT}, we easily obtain
\ba
\< h_{ij}(\vk,\eta) h_{kl}(\vk',\eta') \> =\:& (2\pi)^3 \d_D(\vk-\vk') \\
& \times\left[ e^+_{ij}(\hat\vk) e^+_{kl}(\hat\vk) + e^\times_{ij}(\hat\vk) e^\times_{kl}(\hat\vk) \right] \vs
& \times \frac14 P_T(k,\eta,\eta') \vs
\< h_{ij}(\vk,\eta) h^{ij}(\vk',\eta') \> =\:& (2\pi)^3 \d_D(\vk-\vk') P_T(k,\eta,\eta'). \nonumber
\ea
Long after recombination, the transverse anisotropic stress which sources 
gravitational waves becomes negligible, and the tensor modes propagate
as free waves.  During matter-domination, the tensor 
transfer function then simply becomes
\be
T_T(k,\eta) = 3 \frac{j_1(k \eta)}{k\eta},
\label{eq:TT}
\ee
which however is still valid to a high degree of accuracy during the current 
epoch of acceleration.  We will use \refeq{TT} throughout.  
We also define
\be
\P^{ij} = \d^{ij} - \nhat^i\nhat^j
\ee
as the projection operator perpendicular to the line of sight.  

As a traceless 2-tensor on the sphere, the shear can be decomposed into
spin$\pm2$ functions ${}_{\pm2}\g = \g_1 \pm i \g_2$
(in analogy to the combination of 
Stokes parameters $Q \pm i U$ for radiation) as 
\ba
\g_{ij} =\:& {}_2\g\, m_+^i m_+^j + {}_{-2}\g\, m_-^i m_-^j.
\label{eq:sheardecomp}
\ea
Here, we have defined the unit vectors of the circularly polarized basis,
$\v{m}_\pm \equiv (\v{e}_\theta\mp i\,\v{e}_\phi)/\sqrt{2}$
(see App.~A in \cite{stdruler}).  
If we choose a coordinate system where $\nhat^i$ is along the $z$-axis
and $e_\theta^i$ along the $x$-axis, we have
\be
\g_{ij} = \left(\begin{array}{ccc}
\g_1 & \g_2  & 0 \\
\g_2 & -\g_1 & 0 \\
0 & 0 & 0
\end{array}\right).
\ee
This decomposition is particularly
useful for deriving multipole coefficients and angular power spectra.  
In particular, we can define the multipole moments of the parity-even
$E$-modes and parity-odd $B$-modes through
\ba
a^{\g E}_{lm} =\:& \frac12\left(a^\g_{lm} + a^{\g*}_{lm} \right) \vs
a^{\g B}_{lm} =\:& \frac1{2i}\left(a^\g_{lm} - a^{\g*}_{lm} \right).
\ea 
We also define
\ba
X_\pm \equiv\:& m_\mp^i X_i \vs
E_\pm \equiv\:& m_\mp^i m_\mp^j E_{ij}
\label{eq:Xpm}
\ea
for any vector $X_i$ and tensor $E_{ij}$.  

%%%%%%%%%%%%%%%%%%%%%%%%%%%%%%%%%%%%%%%%%%%%%%%%%%%%%%%%%%%%%%%%%%%%%
%%%%%%%%%%%%%%%%%%%%%%%%%%%%%%%%%%%%%%%%%%%%%%%%%%%%%%%%%%%%%%%%%%%%%
\section{Lensing effects}
\label{sec:lensing}

We begin with the definition of shear in the relativistic context
as derived in \cite{stdruler}, for a synchronous comoving metric
as in \refeq{metric}:
\ba
{}_{\pm2}\g \equiv\:& m_\mp^i m_\mp^j \A_{ij} \vs
=\:& - \frac12 h_\pm - m_\mp^i m_\mp^j \partial_{\perp\,i}\D x_{\perp\,j}.  
\label{eq:shear1}
\ea
In other words, the shear is the traceless part of the symmetric $2\times2$ 
matrix $\A_{ij}$ which describes the transverse distortion of transverse 
standard rulers.  
$\Delta x_\perp^i$ is the displacement perpendicular to the line of sight
of the observed position from the true position of the source, in terms
of the global comoving coordinates.  

As shown in \cite{stdruler}, \refeq{shear1} is explicitly given by
\ba
& ({}_{\pm2}\g)(\vnhat) = -\frac{1}{2} h_{\pm\,o} - \frac12 h_{\pm}
\label{eq:shear11}\\
& \quad - \int_0^{\chit} d\chi
\Bigg\{
\frac{\chit-\chi}{2}\frac{\chi}{\chit}
(m_\mp^i m_\mp^j \partial_i \partial_j h_{kl}) \nhat^k\nhat^l
\vs
& \hspace*{2cm}
+ \left(1-2\frac{\chi}{\chit}\right) \nhat^l m_\mp^k m_\mp^i \partial_i h_{kl}
- \frac1{\chit} h_\pm
\Bigg\}
.\nonumber
\ea
The second term on the r.h.s. can be understood as
coming from the transformation from global coordinates to the local
Fermi normal coordinates \cite{stdruler}; in the following we will
refer to this as the ``FNC term''.    
It is immediately clear that a constant metric perturbation $h_{ij}$
(which corresponds to a pure gauge mode) does not contribute to
$\g_1 \pm i\g_2$.  The same is true for a pure gradient $h_{ij} = B_{ijk} x^k$.  
In App.~C of \cite{stdruler} we have applied several test cases 
to \refeq{shear11},
including a Bianch~I cosmology where all terms contribute non-trivially.  

%%%%%%%%%%%%%%%%%%%%%%%%%%%%%%%%%%%%%%%%%%%%%%%%%%%%%%%%%%%%%%%%%%%%%%%%%%%
%%%%%%%%%%%%%%%%%%%%%%%%%%%%%%%%%%%%%%%%%%%%%%%%%%%%%%%%%%%%%%%%%%%%%%%%%%%
\section{Intrinsic effects}
\label{sec:IA}

\refeq{shear11} is derived assuming we have a perfect standard ruler,
in the sense that the intrinsic physical size of the ruler is 
uncorrelated with the perturbations $h_{ij}$.  In the case of 
weak lensing shear surveys, the ``standard ruler'' is the fact that
galaxies are randomly oriented, i.e. their apparent size measured along two
different fixed directions is on average equal.  However, we know
that in reality galaxy orientations are not truly random, but there is
some alignment with large-scale tidal fields.  

In order to determine the effective tidal field experienced by galaxies
in a Universe with propagating tensor modes, we derive the corrections
$\propto x_{F}^2$ to the metric $g^{F}_{\mu\nu}$ in the Fermi normal
coordinate frame.  In particular, since we are concerned with non-relativistic
motions, we are mostly interested in the time-time
part of the metric $g^{F}_{00}$.  The detailed derivation 
for a space-time described by \refeq{metric} is presented in \refapp{Fermi}.  
The result for $g^{F}_{00}$ is
\ba
g^{F}_{00}(\vx_{F}, t_{F}) =\:& -1 + (\dot H + H^2) r_{F}^2
- 2 \Psi^{F}(\vx_{F}, t_{F}) 
\label{eq:g00F}\\
\Psi^{F}(\vx_F, t_F) =\:& -\frac14 [\ddot h_{ij}(\v{0}, t_{F}) 
+ 2H(t_{F})\dot h_{ij}(\v{0}, t_{F})] x_{F}^i x_{F}^j,
\label{eq:tidal}
\ea
where $r_{F}^2 = \d_{ij} x_{F}^i x_{F}^j$, and dots indicate derivatives
with respect to time $t$ (equivalent to $t_F$ at this order).    
The terms $\propto r_{F}^2$ in \refeq{g00F} are the usual Hubble drag which 
provide an effective repulsive force.  
The leading effect of large-scale cosmological perturbations
on the region considered is to add an effective
tidal field $\Psi^{F}$, which depends on the time derivatives of the
metric perturbation $h_{ij}$.  Note that for a traceless tensor perturbation,
$\Psi^{F}$ is indeed a purely tidal field (i.e. $\nabla^2\Psi^{F} = 0$).  
The fact that the amplitude of the tidal field is given by the 
time derivatives of the tensor modes implies that only modes 
that have entered the horizon contribute;  specifically, the contribution
for a given Fourier mode scales as $k^2$ for $k\to 0$, as can be seen 
from \refeq{TT} and \refeq{shearIA} below.  

In the absence of perturbations ($h_{ij}=0$), \refeq{g00F} is isotropic.  
Thus, in the Fermi frame, there is no preferred direction along which 
galaxies forming in this region could align, and their orientations
are truly random in this case.  As shown in \cite{stdruler}, the expression for
the shear \refeq{shear1} contains precisely this statement (of course
this holds not only for galaxy orientations, but any standard ruler).  

In the presence of perturbations, the tidal field $\propto x_{F}^i x_{F}^j$
provides a preferred direction along which galaxies can align.  
The fact that large-scale tidal fields tend to align galaxies 
(intrinsic alignment, IA) is
well established both theoretically and observationally for
scalar perturbations (e.g., \cite{CatelanKamionkowskiBlandford,HirataSeljak04,BrownEtal02}).  In order to make progress, we will 
adopt the linear alignment (LA) model, which 
has recently been shown to be consistent with observations on large
scales ($\gtrsim 10\Mpch$) \cite{BlazekEtal,JoachimiEtal}.  In this
model (following the notation of \cite{BlazekEtal}), 
the tidal tensor $t_{ij}$ at the location of the galaxy, defined 
through
\be
t_{ij} = \left(\partial_i\partial_j -\frac13 \d_{ij}\nabla^2\right) \Psi,
\ee
contributes to the traceless part of the observed distortion matrix
$\A_{ij}$ of the galaxy image through
\ba
\A^{\rm IA}_{ij}(\vnhat) =\:& -\frac{C_1}{4\pi G} \P_{ik} \P_{jl} t^{kl}(z_p) \vs
=\:& -\frac23 C_1 \rhocr H_0^{-2} \P_{ik} \P_{jl}\, t^{kl}(z_p).
\label{eq:LA}
\ea
That is, $\A^{\rm IA}_{ij}$ is proportional to the projection of the tidal
tensor onto the sky plane.  Here, $\rhocr = 3 H_0^2/(8\pi G)$ 
is the critical density 
today.  The constant of proportionality $C_1$ determines the
magnitude of alignment, while $z_p$, the redshift at which the tidal
field is evaluated, is another parameter of the model.  
Observationally, $C_1\rhocr \sim 0.1$ for galaxies at redshifts less than 1,
when choosing $z_p$ to be equal to the source redshift.  
A positive $C_1$, together
with the overall sign in \refeq{LA}, corresponds to a galaxy's major axis
aligning with
the \emph{smallest} eigenvector of $t_{ij}$; physically, an initially
spherical perturbation will tend to collapse last in the direction of
the smallest potential curvature, leading to a preferential alignment
of the major axis with this direction.  Note that in \refeq{LA},
$C_1\rhocr$ multiplies the tidal field in physical rather than
comoving units.  

While this mechanism is expected to be qualitatively the same for
tensor modes as for scalar modes, there is no reason for the amplitude $C_1$
to be the same in both cases.  In particular, linear tidal fields sourced
by scalar perturbations are constant during matter domination, while
tensor perturbations decay and oscillate.  However, we will see
that the tensor modes relevant for the IA contribution are long-wavelength
and should not have a qualitatively different impact 
on the formation of galaxies than scalar tidal fields.  We will return
to this issue in \refsec{res}.  In the following,
we will assume that $z_p$ is equal to the source redshift.  Generally, 
evaluating the tidal field at $z_p > \zt$ leads to larger effects so that 
this is a conservative assumption.

Using \refeq{tidal} and \refeq{shear1}, it
is then straightforward to evaluate the contribution of tensor modes
to the shear ${}_{\pm 2} \g$:
\ba
({}_{\pm 2} \g)^{\rm IA}(\vnhat) =\:& m_\mp^i m_\mp^j \A^{\rm IA}_{ij} \vs
=\:& \frac13 C_1\rhocr H_0^{-2} 
\left(\ddot h_{\pm} + 2 H \dot h_{\pm}\right) \vs
=\:& \frac13 C_1\rhocr H_0^{-2} a^{-2}
\left(h_\pm'' + a H h_\pm'\right),
\label{eq:shearIA}
\ea
where primes indicate derivatives with respect to $\eta$, and we have
used $dt = a d\eta$.  \\

%%%%%%%%%%%%%%%%%%%%%%%%%%%%%%%%%%%%%%%%%%%%%%%%%%%%%%%%%%%%%%%%%%%%%
%%%%%%%%%%%%%%%%%%%%%%%%%%%%%%%%%%%%%%%%%%%%%%%%%%%%%%%%%%%%%%%%%%%%%
\section{Observed shear statistics}
\label{sec:Clshear}

We now use the results derived in the previous two sections to
calculate the angular power spectrum of the observed shear
induced by tensor modes.  We briefly outline the steps of 
the derivation, which follows the general prescription described
in App.~A1 of \cite{stdruler}, with details relegated to \refapp{shear}.

In the first step, we consider a single tensor mode of wavevector
$\vk$ which we assume oriented along the $z$-axis.  Including
the intrinsic alignment contribution (\refsec{IA}), we
obtain for the contribution to the shear
\begin{widetext}
\ba
 ({}_{\pm2}\g)(\vk,\vnhat) = \sum_{p=-1,1}\Bigg\{ &
- \frac12 \left[ h_p(\vk,\eta_0) + 
\left(1 - \frac23 C_1\rhocr H_0^{-2}\tilde a^{-2} \left\{\partial_{\tilde\eta}^2 + \tilde a \tilde H \partial_{\tilde\eta}\right\}\right)
h_p(\vk,\tilde\eta) e^{i\vk\cdot\vnhat\,\chit}\right] \frac12 (1\mp p\mu)^2 e^{i2p\phi} \vs
 & + \int_0^{\chit} d\chi
\bigg[
\frac{\chit-\chi}{4}\frac{\chi}{\chit}
k^2 (1-\mu^2)^2
+ \left(1-2\frac{\chi}{\chit}\right) 
i \frac{k}2 (1-\mu^2) (1\mp p\mu)
+ \frac1{2\chit}(1\mp p\mu)^2 
\bigg] \vs
& \hspace*{1.2cm}
\times e^{i2p\phi} h_p(\vk,\eta_0-\chi) e^{i\vk\cdot\vnhat\,\chi}\Bigg\},
\nonumber
\ea
where $h_{\pm1} = (h_+ \mp i h_\times)/2$ are the circular polarization
states of tensor modes, $\mu = \cos\theta$, and $\theta, \phi$ denote 
the polar and azimuthal angles of the line of sight unit vector $\vnhat$.  
We then apply the spin-lowering/raising operator twice
to ${}_{\pm2}\g$ in order to obtain a rotationally invariant
(spin-0) quantity.  The spherical harmonic coefficients of this
scalar quantity are directly related to those of the spin$\pm2$ shear
components (see App.~A in \cite{stdruler}).  
We can then separate $a^\g_{lm}$ into a real part $a^{\g E}_{lm}$, which 
is parity-even and thus transforms in the same way as a tensor derived from a
scalar function $f$, and an imaginary part
$i\,a^{\g B}_{lm}$, which acquires an additional minus sign under
parity.  This $E/B$-mode decomposition is useful because any 
symmetric tensor $\g_{ij}$ derived from a scalar function is parity-even
and thus does not source any $B$-modes (as shown explicitly for the shear
in \cite{stdruler}).  
Further, any perturbations generated by parity-conserving physics do no
induce an $E-B$ cross-correlation.  

Finally, using relations derived in App.~A1 of \cite{stdruler}, we obtain
the angular power spectra of $E$- and $B$-modes of the shear induced
by tensor modes (\refapp{shear2}):
\ba
C_\g^{XX}(l) =\:& \frac1{2\pi} 
\int k^2 dk\: P_{T0}(k) |F_l^{\g X}(k)|^2,\qquad X = E,\,B \label{eq:Clshear1} \\
%%%
F_l^{\g E}(k) \equiv\:& 
- \frac14 
\left[ T_T(k,\eta_0) \left(\Re \hat Q_1(x) \frac{j_l(x)}{x^2}\right)_{x=0}
+ \left(1 - \frac23 C_1\rhocr H_0^{-2}\tilde a^{-2} \left\{\partial_{\tilde\eta}^2 + \tilde a \tilde H \partial_{\tilde\eta}\right\}\right)
T_T(k,\tilde\eta)\Re \hat Q_1(\tilde x) \frac{j_l(\tilde x)}{\tilde x^2}\right] \vs
 & + \int_0^{\chit} \frac{d\chi}\chi
\bigg[ \Re \hat Q_2(x) + \frac{\chi}{\chit} \Re \hat Q_3(x)
\bigg]
\frac{j_l(x)}{x^2} T_T(k, \eta_0-\chi)
\vs
%%%
F_l^{\g B}(k) \equiv\:& 
- \frac14 
\left[ T_T(k,\eta_0) \left(\Im \hat Q_1(x) \frac{j_l(x)}{x^2}\right)_{x=0}
+ \left(1 - \frac23 C_1\rhocr H_0^{-2}\tilde a^{-2} \left\{\partial_{\tilde\eta}^2 + \tilde a \tilde H \partial_{\tilde\eta}\right\}\right)
T_T(k,\tilde\eta)\Im \hat Q_1(\tilde x) \frac{j_l(\tilde x)}{\tilde x^2}\right] \vs
 & + \int_0^{\chit} \frac{d\chi}\chi
\Im\hat Q_2(x) \frac{j_l(x)}{x^2} T_T(k, \eta_0-\chi).
\nonumber
\ea
\end{widetext}
Here $\hat Q_i(x)$ are derivative operators whose action on $j_l(x)/x^2$
is given explicitly in \refeq{Qibessel}; in particular $\Im \hat Q_3(x)[j_l(x)/x^2] = 0$.    

% % % % % % % % % % % % % % % % % % % % % % % % % % % % % % % % % % % % % 
\subsection{Relation to convergence and rotation}
\label{sec:kom}

The shear is most commonly written in terms of angular derivatives
of the deflection angle $\Delta\theta^i$, i.e. $\g_{ij}$ is defined as the
trace-free part of $\partial \Delta\theta_j/\partial \theta_i$.  In our notation,
$\Delta\theta^i = \Delta x_\perp^i/\chit$, and 
$\partial \Delta\theta_j/\partial \theta_i=\partial_{\perp i} \D x_{\perp j}$.  There are two degrees of
freedom in $\D x_{\perp}^i$, and hence only two independent components
of $\partial_{\perp i} \D x_{\perp j}$.  If we define the convergence
$\hat\k$ and rotation $\omega$ through
\ba
\hat\k \equiv\:& -\frac12 \partial_{\perp i} \D x_\perp^i \vs
\omega \equiv\:& -\frac12 \eps_{ijk} \nhat^i\partial_\perp^j \Delta x_\perp^k,
\label{eq:kom}
\ea
then the $E$-mode of the shear defined as trace-free part of
$\partial_{\perp i}\D x_{\perp j}$ is directly related to $\hat\k$, while
the $B$-mode is related to $\omega$.  As shown by \citet{Stebbins},
on the full sky
\ba
C^{EE}_\g(l) =\:& \frac{(l+2)(l-1)}{l(l+1)} C_{\hat\k}(l) \vs
C^{BB}_\g(l) =\:& \frac{(l+2)(l-1)}{l(l+1)} C_{\omega}(l).
\label{eq:CgCk}
\ea
However, neither of the deflection angle $\D x_\perp^i$, the
convergence $\hat\k$, or the shear defined as trace-free part of
$\partial_{\perp i}\D x_{\perp j}$ are observable (the rotation is only
observable if there is an intrinsic preferred direction in the source
plane).  Instead,
the observable shear is given by \refeq{shear11} which includes the 
FNC term.  Nevertheless, the relations \refeq{CgCk} are useful as an analytical
and numerical cross-check of \refeq{Clshear1} (without the IA contribution), 
and this cross-check is presented in \refapp{kom}.  

In the previous work of \cite{DodelsonEtal}, the rotation was used
as a proxy for shear $B$-modes through \refeq{CgCk}.  This result thus
does not include the FNC term.  However, they also consider the
``metric shear'', which is the contribution to $\omega$ that
corresponds to the shear contribution by the FNC term.  When including
the metric shear in $\omega$, we indeed recover the $B$-modes of the
shear including the FNC term through the relation \refeq{CgCk}
(see \refapp{rot}).  Thus, our results for the lensing-induced 
shear $B$-modes from tensor modes (i.e., neglecting the intrinsic 
alignment contribution) agree with those of \cite{DodelsonEtal}, modulo
the factor $(l+2)(l-1)\,/\,l(l+1)$.  

%!!!!!!!!!!!!!!!!!!!!!!!!!!!!!!!!!!!!!!!!!!!
\begin{figure}[t!]
\centering
\includegraphics[width=0.5\textwidth]{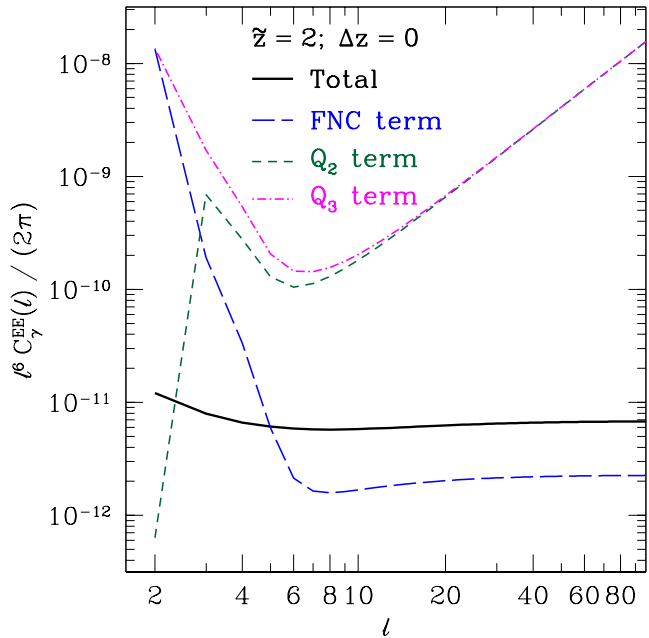}
\caption{Lensing (projection) contributions to the observed angular 
power spectrum of
the $E$-mode component of the shear from tensor modes, separated
into terms $\propto \hat Q_1$ (observer and FNC terms), $\propto \hat Q_2$
and $\propto \hat Q_3$ respectively.  Note that power spectra 
are multiplied by $l^6$.  
We have assumed a sharp source redshift of $\tilde z = 2$.  
}
\label{fig:Cl_E}
\end{figure}
%!!!!!!!!!!!!!!!!!!!!!!!!!!!!!!!!!!!!!!!!!!!

%%%%%%%%%%%%%%%%%%%%%%%%%%%%%%%%%%%%%%%%%%%%%%%%%%%%%%%%%%%%%%%%%%%%%
%%%%%%%%%%%%%%%%%%%%%%%%%%%%%%%%%%%%%%%%%%%%%%%%%%%%%%%%%%%%%%%%%%%%%
\section{Results}
\label{sec:res}

\begin{figure}[t!]
\centering
\includegraphics[width=0.5\textwidth]{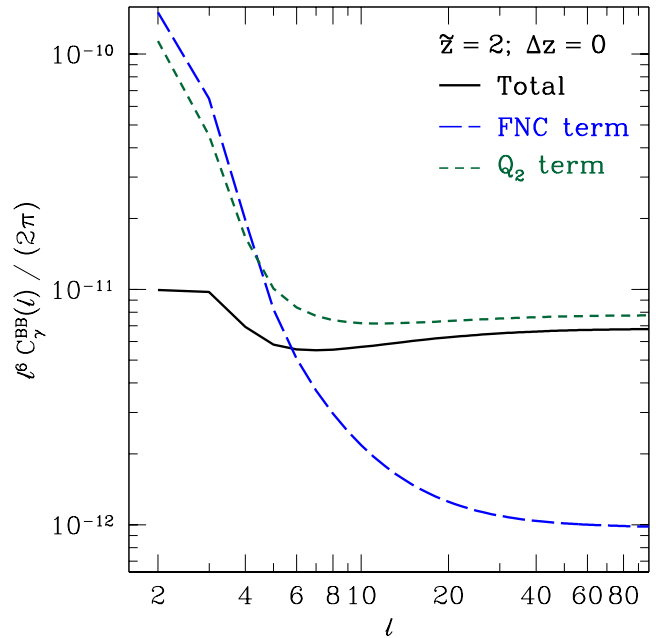}
\caption{Same as \reffig{Cl_B}, but for the $B$-mode component.
}
\label{fig:Cl_B}
\end{figure}

We begin by investigating the separate terms contributing to
the shear $E$- and $B$-mode power spectra in \refeq{Clshear1}, focusing
on the lensing contributions without intrinsic alignment first.  
\reffig{Cl_E} shows the $E$-mode angular power spectra obtained by separately
considering only the terms $\propto \hat Q_2,\,\propto \hat Q_3$,
and the FNC and observer terms $\propto \hat Q_1$.  Here, we have
multiplied the power spectra by $l^6$ as they are very steeply falling with
$l$.   We see that 
for $l \lesssim 10$ there is a very significant cancellation
(by three orders of magnitude) between the different terms.  In fact,
the magnitude of the individual terms depends on the lower limit used
in the integration over $k$, and diverges logarithmically as $k_{\rm min}\to 0$
for small $l$.  
For the $B$-modes, the cancellation is not as important, though
still significant (\reffig{Cl_B}; in this case there is no
$\hat Q_3$ term).  This result is in agreement with the findings of
\cite{DodelsonEtal}:  when including the FNC contribution (metric shear),
the amplitude decreases significantly.  This is not surprising: if we
drop the FNC term in \refeq{Clshear1}, tensor perturbations with $k\to 0$
contribute to the shear at low $l$, and since the tensor power spectrum
is sharply falling with $k$, these contributions are large.  Clearly,
these contributions are unphysical however, and the FNC term must be included.  

\begin{figure}[t!]
\centering
\includegraphics[width=0.5\textwidth]{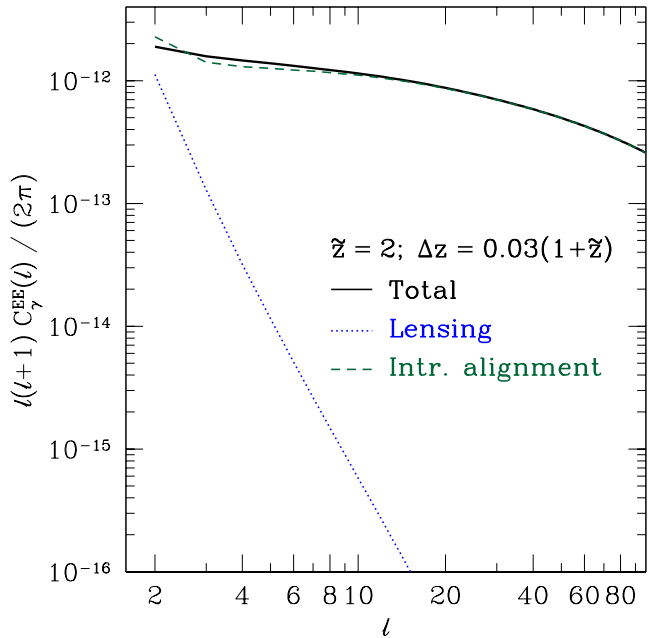}
\caption{Angular power spectrum of the observed $E$-component
of the shear from lensing and intrinsic alignment affects,
as well as the total power spectrum.  We assumed
$C_1\rhocr = 0.12$ (following the results of \cite{BlazekEtal}),
and a Gaussian distribution of source redshifts centered at
$\zt = 2$ with RMS width of $\D z = 0.03(1+\zt)$.  
}
\label{fig:Cl_IAE}
\end{figure}

\begin{figure}[t!]
\centering
\includegraphics[width=0.5\textwidth]{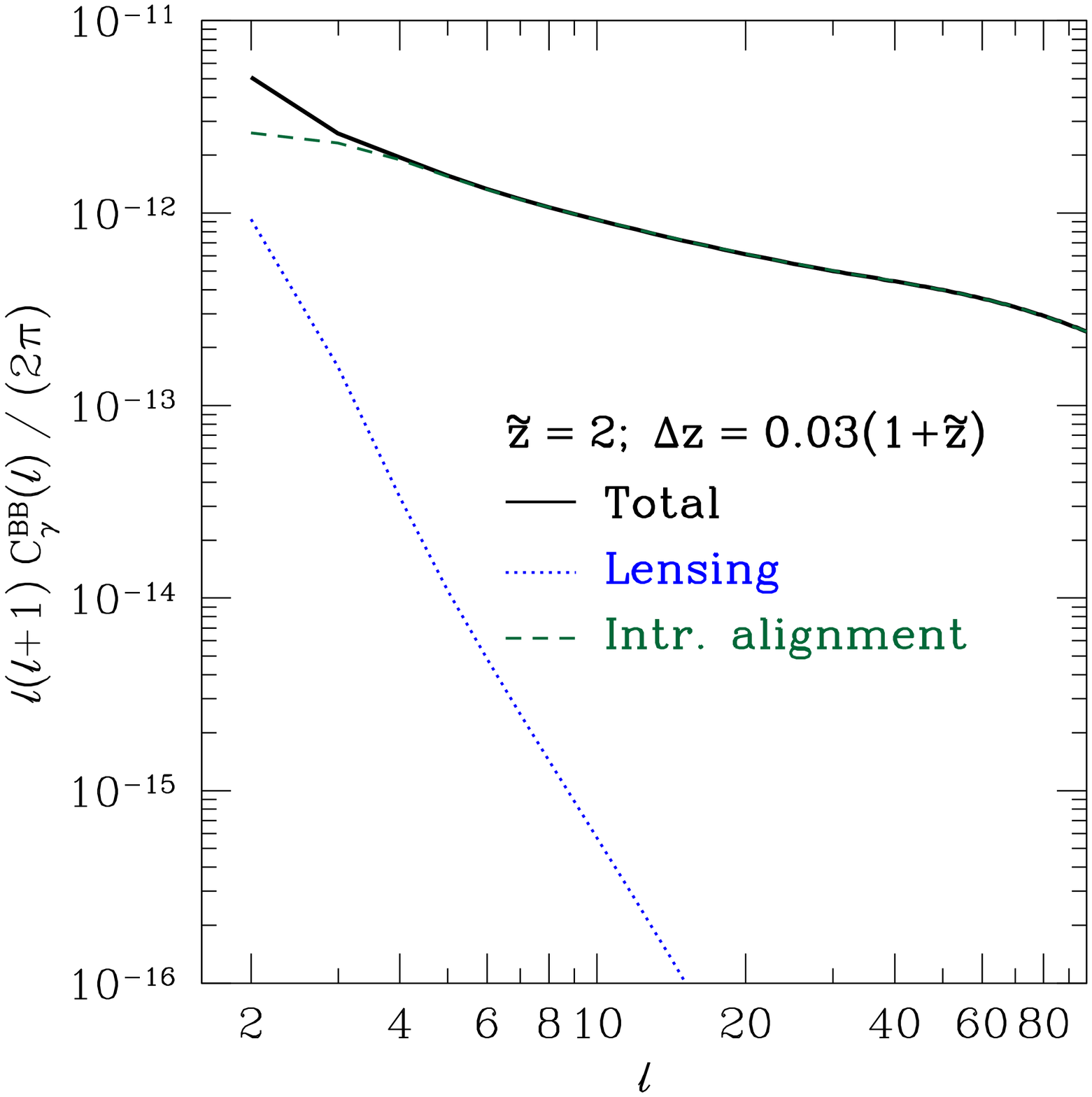}
\caption{Same as \reffig{Cl_IAE}, but for $B$-modes.
}
\label{fig:Cl_IAB}
\end{figure}

\reffig{Cl_IAE} shows the total lensing contribution to the $E$-mode
power spectrum of the shear, the intrinsic alignment contribution,
as well as the sum of the two, while \reffig{Cl_IAB} shows the same
for the (more interesting) $B$-modes.  Here, we assumed a Gaussian
redshift distribution $dN/d\zt\propto \exp(-(\zt-\bar z)^2/2\D z^2)$
with $\D z = 0.03(1+\bar z)$.  Further,  we have adopted a value of 
$C_1\rhocr = 0.12$ as measured in the Sloan Digital Sky Survey
\cite{BlazekEtal}.  This coefficient will depend on the specific
galaxy sample considered, in particular on the redshift.  Here we extrapolate
the value of $C_1\rho_{\rm cr 0}$ measured for galaxies at 
$z \approx 0.3-0.5$ to galaxies at $z=2$, assuming no evolution.  Thus 
our results should only be seen as a rough estimate of the magnitude of this
effect (note however that we assume a constant alignment strength with
respect to the physical, not comoving tidal field).  
Even with this caveat in mind however, it is clear that the intrinsic
alignment contribution is far larger than the lensing contribution.  
This is in contrast to the scalar case, where for source galaxies
at cosmological redshifts the lensing signal is significantly larger
than the intrinsic alignment contribution.  The underlying reason is 
that the projected contributions from lensing are relatively suppressed
in the tensor case.  While scalar perturbations with transverse wavevector
deflect light coherently along the past line cone to the source, 
tensor perturbations propagate and decay, such that no such coherent
deflection occurs even for transverse wavevectors \cite{KaiserJaffe}.  
The result is that the lensing contributions are mostly localized at the source
for tensor modes,
and down-weighted by the lensing kernel ($\propto \chit-\chi$).  

%!!!!!!!!!!!!!!!!!!!!!!!!!!!!!!!!!!!!!!!!!!!
%!!!!!!!!!!!!!!!!!!!!!!!!!!!!!!!!!!!!!!!!!!!

Apart from $C_1\rhocr$, the linear alignment model has another free parameter
in the redshift $z_p$ at which the tidal field is evaluated.  By default,
we choose $z_p = \zt$.  However, choosing $z_p$ to correspond to a 
time $5\times 10^8$ years before observation ($z_p \approx \zt + 0.04$ for 
$\zt = 2$), which corresponds to 
several dynamical times for typical galaxies, only yields a slight increase 
in the power spectrum contribution by $\sim$3\%.  Varying $z_p$ thus does
not have a significant impact on the intrinsic alignment contribution.  
On the other hand, this mild dependence on $z_p$ indicates that the bulk
of the IA contribution induced by tensor modes is due to slowly varying
tidal fields, i.e. tensor modes with $k/H \sim 1$, rather than rapidly
oscillating modes with $k/H \gg 1$ (this is confirmed by numerical
inspection of the intrinsic alignment contribution to $F_l^{E,B}$).  
Such tidal fields, which vary on
a Hubble time, are not expected to have a qualitatively different effect
on the formation of galaxies and halos than scalar tidal fields, given that
the relevant time scale is the dynamical time of the collapsing dark matter halo.  
We thus expect that
the value of $C_1\rhocr$ relevant for the IA contributions to shear
from tensor modes will not be very different from that for scalar tidal
fields.  However, one would expect $C_1\rhocr$ to be generically 
scale-dependent for tensor modes, decaying from its low-$k$ limit to
smaller values as $1/k$ approaches the scale of halos and galaxies 
($k \gtrsim 0.3 \iMpch$).  

%!!!!!!!!!!!!!!!!!!!!!!!!!!!!!!!!!!!!!!!!!!!
%!!!!!!!!!!!!!!!!!!!!!!!!!!!!!!!!!!!!!!!!!!!

\begin{figure}[t!]
\centering
\includegraphics[width=0.5\textwidth]{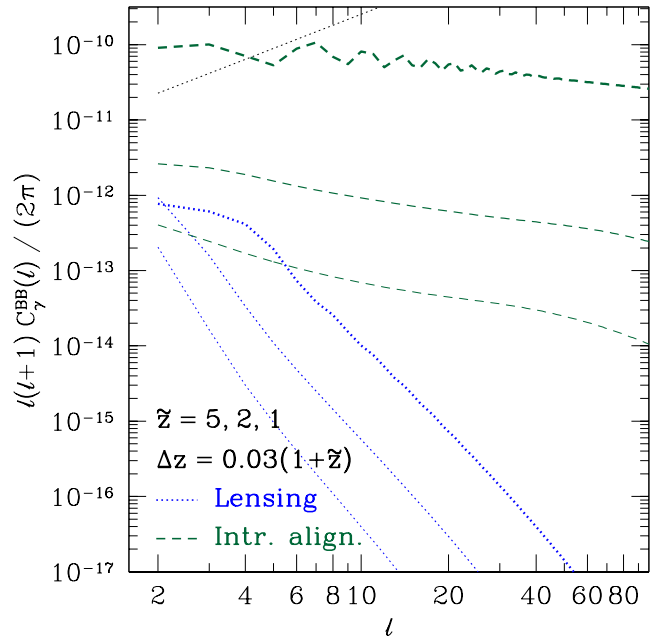}
\caption{Dependence of the lensing and intrinsic alignment contributions
to the $B$-mode shear power spectrum on the source redshift $\tilde z$.  
We have assumed a Gaussian redshift distribution centered at
$\zt = 5$, 2, 1 (from top to bottom), and RMS width
$\D z = 0.03 (1+\zt)$.  The black dotted
line near the top of the figure shows the $1\sigma$ error on the
shear power spectrum per multipole induced by shape noise [\refeq{Clshape}], 
for a survey with $\bar n = 100\:{\rm arcmin}^{-2}$, $\sigma_e = 0.3$, and $\fsky = 0.5$.  
}
\label{fig:Cl_z}
\end{figure}

\begin{figure}[t!]
\centering
\includegraphics[width=0.5\textwidth]{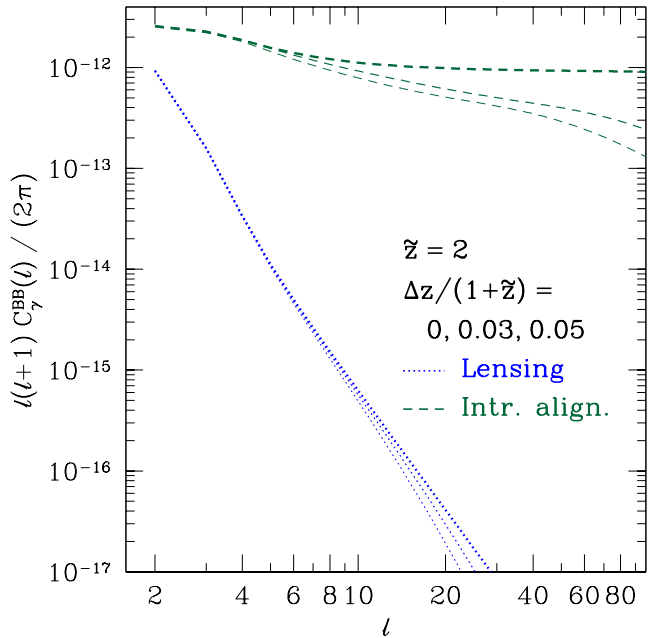}
\caption{Dependence of the lensing and intrinsic alignment contributions
to the $B$-mode shear power spectrum on the width of the source redshift 
distribution $\D z$.  $\tilde z = 2$ for all curves.  
}
\label{fig:Cl_Dz}
\end{figure}

\reffig{Cl_z} shows the redshift evolution of the $B$-modes of the shear.  
As expected, larger source redshifts yield significantly larger signals,
due to the decay of the tensor modes and
since at higher redshifts, larger scales are being probed at a given $l$.  
However, we also see that the intrinsic alignment contribution evolves
even faster with source redshift (note that here we have assumed the same
value for $C_1\rhocr$ at all redshifts).  This can be traced back to
the factor of $\tilde a^{-2}$ in the IA contribution [\refeq{Clshear1}],
which is due to the transformation from conformal time derivatives to
physical time derivatives.  It is also interesting to 
consider the dependence of the signal on the width of the source
galaxy redshift distribution.  This is illustrated in \reffig{Cl_Dz}.  
The lensing contributions are largely independent of $\D z$ 
for the range of multipoles relevant here.  On the other hand, 
the IA contribution is noticeably increased for sharp source redshifts
at $l \gtrsim 10$,
a consequence of the fact that this contribution is not projected
along the line of sight but evaluated at the source.  Thus, unlike the
lensing contribution, the IA contribution is essentially a three-dimensional 
field.  Note also
that in this case $l(l+1) C^{BB}_\g(l) \approx$~const, i.e. there
is roughly equal power per decade in multipole for the IA contribution.    
However, following our discussion above, we expect the approximation of a
scale-independent alignment coefficient to break down once the 
wavelength of contributing tensor modes approaches the scale of halos, 
roughly at $l$ greater than a few hundred.  

%!!!!!!!!!!!!!!!!!!!!!!!!!!!!!!!!!!!!!!!!!!!
%!!!!!!!!!!!!!!!!!!!!!!!!!!!!!!!!!!!!!!!!!!!

\subsection{$B$-modes from scalar perturbations}
\label{sec:scalar}

In addition to being sourced by primordial tensor modes, 
shear $B$-modes are also produced by
second-order corrections to lensing by scalar perturbations (here
we neglect second order corrections to the scalar intrinsic alignment
contributions).  These come from three sources.  

The first source
is tensor modes generated by non-linear gravitational instability 
(e.g., \cite{MollerachEtal04,BaumannEtal07}).  The shear $B$-mode 
power spectrum induced by these tensor modes was found to be roughly 
scale-invariant and at the level of $l(l+1) C^{BB}(l)/2\pi \approx 10^{-14}$ 
\cite{SarkarEtal08} which is much smaller than the total primordial GW signal 
presented here for $r=0.2$.  However, the calculation of \cite{SarkarEtal08}
did not include the intrinsic alignment effect which will also increase
the $B$-mode signal of scalar-induced tensor modes.  Due to the different
scale-dependence and redshift evolution of the latter, the boost will likely
be somewhat smaller than that for the primordial tensor modes.  We
leave this for future work.  

\begin{figure}[t!]
\centering
\includegraphics[width=0.5\textwidth]{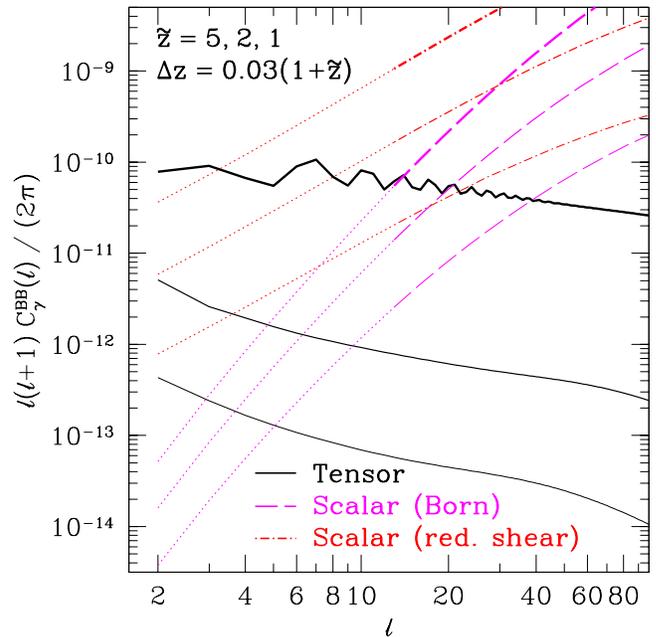}
\caption{Shear $B$-modes from tensor perturbations (black solid, including
both lensing and intrinsic alignment) and second order scalar contributions
from corrections to the Born approximation (magenta dashed) and
reduced shear and lensing bias (assuming $q=1$, red dash-dotted).  
Results are for redshifts $\zt = 5$, 2, 1 (from top to bottom).  Note
that the scalar contributions have been calculated using the Limber 
and flat-sky approximations, respectively, which are not accurate
at $l \lesssim 10$ (dotted lines).
}
\label{fig:Cl_2nd}
\end{figure}

The second, more significant scalar source for $B$-modes is from second-order
corrections in the geodesic equation (beyond-Born correction and lens-lens
coupling) \cite{CoorayHu02,HirataSeljak03}.  We have evaluated these
according to 
\ba
C_\gamma^{BB} =\:& \frac{(l+2)(l-1)}{l(l+1)} C_l^{\omega\omega},
\ea
where $C_l^{\omega\omega}$ is given in Eq.~(51) of \cite{HirataSeljak03}.  
Note that the expression given there uses the Limber and flat-sky
approximations and is thus not expected to be accurate for $l \lesssim 10$.  
We compare this contribution with the total contribution from primordial
tensor modes (intrinsic alignment and lensing) in \reffig{Cl_2nd}.  
At $z=2$, the scalar contributions become larger than the tensor mode signal 
at $l \gtrsim 6$.  For higher redshifts, they only dominate at higher $l$.  

Finally, a third contribution comes from the fact that observationally
we measure the reduced shear $g = \gamma/(1-\kappa)$ \cite{Schneider:1997ge,Dodelson:2005ir}.  Furthermore,
selection effects (``lensing bias'') lead to a similar second order 
correction \cite{sizebias}.  At leading order, both contributions
can be summarized by writing the observed shear tensor as
\be
\g^{\rm obs}_{ij}(\vnhat) = \g_{ij}(\vnhat) + (1+q) \hat\k(\vnhat)\g_{ij}(\vnhat),
\ee
where the parameter $q$ parametrizes the lensing bias contribution \cite{sizebias}.  
We can evaluate the scalar $B$-mode contribution 
from both of these effects using Eq.~(21) in \cite{sizebias}.  Note that
this equation was derived in the flat-sky limit and hence will also not be
accurate at $l \lesssim 10$.  This contribution is also shown in \reffig{Cl_2nd} 
(assuming a lensing bias coefficient of $q=1$).  This contribution is even 
larger than that from the second order Born correction, and in fact
dominates over the primordial GW contribution at $\zt=2$ for all $l$ but 
$l=2$.  In principle one could reduce this contribution significantly by selecting
a source galaxy sample with $q \approx -1$, although whether this is
feasible in practice would need to be investigated.  The fact that 
the reduced shear and lensing bias contributions produce the dominant 
scalar contribution to shear $B$-modes is an interesting result in itself
and has not been pointed out before.

%!!!!!!!!!!!!!!!!!!!!!!!!!!!!!!!!!!!!!!!!!!!
%!!!!!!!!!!!!!!!!!!!!!!!!!!!!!!!!!!!!!!!!!!!

%!!!!!!!!!!!!!!!!!!!!!!!!!!!!!!!!!!!!!!!!!!!
%!!!!!!!!!!!!!!!!!!!!!!!!!!!!!!!!!!!!!!!!!!!

%!!!!!!!!!!!!!!!!!!!!!!!!!!!!!!!!!!!!!!!!!!!
%!!!!!!!!!!!!!!!!!!!!!!!!!!!!!!!!!!!!!!!!!!!

%%%%%%%%%%%%%%%%%%%%%%%%%%%%%%%%%%%%%%%%%%%%%%%%%%%%%%%%%%%%%%%%%%%%%
%%%%%%%%%%%%%%%%%%%%%%%%%%%%%%%%%%%%%%%%%%%%%%%%%%%%%%%%%%%%%%%%%%%%%
\section{Discussion}
\label{sec:concl}

In this paper, we have studied the shear induced by a primordial GW background.  
In addition to the projection (lensing) effects, for which we use
a gauge-invariant expression, we derive for the first time the
contribution due to intrinsic alignment (IA) of galaxies through the
effective tidal field induced by tensor modes.  We have found
that this contribution is typically much larger than that from lensing.  
While surprising initially, this is due to the qualitatively different
properties of lensing by tensor modes as compared to scalar modes.  

The IA contribution depends on a coefficient which can be observationally
determined for scalar perturbations.  In general, this does not have
to be the same for tensor modes, since tensor modes evolve while the scalar
tidal field is constant on large scales (during matter domination).  
On the other hand, the bulk of the
tensor contributions is from horizon-scale modes, which evolve on the Hubble
time scale.  Compared to the dynamical time of galaxies and halos, this is
a very slow evolution, and we expect the alignment coefficient to only
be mildly affected by this.  On the other hand, the results shown in
\reffig{Cl_IAE} through \ref{fig:Cl_Dz} depend on the alignment strength
at high redshifts, which is currently poorly known observationally.  

The IA contribution also decays much more
slowly towards high $l$ than the lensing contribution (especially for
narrow source redshift distributions).  In principle, this could allow one
to access smaller-scale tensor modes than those probed by the cosmic microwave 
background.  Further, since the IA contribution is not projected along the 
light cone, this effect in principle allows us to measure the entire
three-dimensional
field of tensor perturbations.  In the case of perfect redshift measurements,
the number of tensor modes measurable up to a maximum scale would thus scale
as $k_{\rm max}^3$ rather than $l_{\rm max}^2$.  

On the other hand, even with this increased signal, the requirements 
for a detection of a stochastic GW background using galaxy ellipticities 
are still extremely challenging.  Purely in terms of statistical power, 
the intrinsic ellipticities
of galaxies add noise (``shape noise'') to the shear $E$- and $B$-mode 
power spectra.  The corresponding 1$\sigma$ error on the power spectrum
per multipole is given by
\ba
\D C^{XX}_\gamma(l) = \frac1{\sqrt{(2l+1) f_{\rm sky}}}\, \frac{\sigma_e^2}{2\bar n},
\ea
where $X=E,B$, $f_{\rm sky}$ is the fraction of sky covered by the survey,
$\sigma_e$ is the RMS intrinsic ellipticity of galaxies, and $\bar n$
is the number of source galaxies per steradian.  In numbers, this yields
\ba
\frac{l(l+1)}{2\pi} \D C^{XX}_\gamma(l) \approx\:& 1.6\times 10^{-11} 
\left(\frac{l}2\right)^{3/2} \fsky^{-1/2} 
\left(\frac{\sigma_e}{0.3}\right)^2  \vs
& \times \left(\frac{\bar n}{100\:\rm arcmin^{-2}}\right)^{-1}.
\label{eq:Clshape}
\ea
This prediction, for $\fsky = 0.5$, is shown as black dotted line
in \reffig{Cl_z}.  Clearly, one would need to go to
source redshifts $\zt > 2$, at very high source densities, to detect
a GW contribution at the level of $r=0.2$ (unless the alignment strength
at $z \sim 2$ is significantly larger than that measured at low redshifts).  

Given the smallness of the signal and the possible contamination by
scalar contributions (\refsec{scalar}), 
quantitative constraints on an inflationary GW 
background will thus be very challenging.  Nevertheless, shear $B$-modes 
remain one of only a
handful of probes in cosmology that can be used to search for such
a background.

%%%%%%%%%%%%%%%%%%%%%%%%%%%%%%%%%%%%%%%%%%%%%%%%%%%%%%%%%%%%%%%%%%%%%%%%%%%
\acknowledgments
We would like to thank Jonathan~Blazek, Chris~Hirata, Marc~Kamionkowski, 
Eiichiro~Komatsu, Samuel Gralla, Uro$\check{\rm s}$ Seljak, and Masahiro~Takada 
for helpful discussions.  
FS is supported by the Gordon and Betty Moore Foundation at Caltech.  
DJ was supported by NASA NNX12AE86G.

%%%%%%%%%%%%%%%%%%%%%%%%%%%%%%%%%%%%%%%%%%%%%%%%%%%%%%%%%%%%%%%%%%%%%%%%%%%
%%%%%%%%%%%%%%%%%%%%%%%%%%%%%%%%%%%%%%%%%%%%%%%%%%%%%%%%%%%%%%%%%%%%%%%%%%%
\appendix

%%%%%%%%%%%%%%%%%%%%%%%%%%%%%%%%%%%%%%%%%%%%%%%%%%%%%%%%%%%%%%%%%%%%%%%%%%%
\begin{widetext}
%\clearpage

%%%%%%%%%%%%%%%%%%%%%%%%%%%%%%%%%%%%%%%%%%%%%%%%%%%%%%%%%%%%%%%%%%%%%%%%%%%
\section{Fermi normal coordinates}
\label{app:Fermi}

In this appendix we review the basic concept of Fermi normal coordinates
and its application to a flat FRW metric with perturbation $h_{ij}$ of the 
spatial components (\refeq{metric}, but without imposing the transverse
or tracefree conditions on $h_{ij}$).  General covariance allows us
to choose coordinates such that, at any given space-time point $P$, 
the metric is Minkowski and the Christoffel symbols vanish \cite{MTW}; i.e.,
\ba
g_{\mu\nu} =\:& \eta_{\mu\nu}
\vs
g_{\mu\nu,\alpha} =\:& 0.
\label{eq:localMinkowskian}
\ea
\emph{Riemann} normal coordinates realize such coordinates by using 
a set of four geodesics (one time-like and three space-like)
starting from the fixed point $P$.  
\emph{Fermi} normal coordinates (FNC) are a specific
extension of Riemann coordinates such that \refeq{localMinkowskian} holds for 
every point along a fixed time-like geodesic.  Specifically, 
we single out a time-like geodesic (``central geodesic'') 
that passes through $P$, as the worldline of the observer around which 
the FNC are constructed.  
Given three space-like tangent vectors
at $P$ which are orthogonal to the tangent vector of the central geodesic 
at $P$, the tangent vectors at all other points along the central geodesic
are defined through parallel transport.  
For all points along the central geodesic, we construct Riemann 
coordinates using these tangent vectors.  Then,
the condition \refeq{localMinkowskian} is satisfied at all points 
along the central geodesic.  
Given a central geodesic, 
FNC are uniquely defined up to three Euler angles.  The 
significance of FNC is that they are the natural coordinates in which
an observer moving along the central geodesic would describe local
experiments.  
With the conditions \refeq{localMinkowskian} satisfied, 
the departure from Minkowski of the metric in FNC appears at 
quadratic order in the spatial Fermi coordinate $x_{F}^i$.  

The Fermi normal coordinates can be explicitly constructed as follows (see \cite{ManasseMisner,BaldaufEtal}).  
The time coordinate is chosen to coincide with the proper time  
along the central geodesic.  
Let $x^0_F(P) = t_P$ denote its value at point $P$.  
We can construct the spatial
slicing of FNC, i.e. the $x^0_F=$~const hypersurface, as comprising
all space-time points in the neighborhood of the central geodesic 
that are reached by a congruence of spatial geodesics whose tangent vector
at $P$ are orthogonal to the tangent vector of the central geodesic
$(e_0)^\mu_P$.  Let $Q$ denote a point on this hypersurface, so that
$x^0_F(Q) = t_P$.  Further let
$x^\mu(\lambda)$ be the unique geodesic (up to reparametrization) 
that connects $P$ and $Q$.  We can fix the
affine parameter by requiring $x^\mu(0)=P,\;x^\mu(1) = Q$.  
We now expand $x^\mu(\lambda)$ in a power series in $\lambda$ around
$\lambda=0$,
\be
x^\mu(\lambda) 
= 
\sum_{n=0}^\infty
\alpha_n^\mu\lambda^n.
\label{eq:Fermiansatz}
\ee
The requirement that $x^\mu(0) = P$ constrains $\alpha_0^\mu$ to 
be equal to the coordinates of $P$ in the chosen, arbitrary coordinate system.  
Given a set of three orthonormal space-like unit vectors $(e_i)^\mu_P$ at $P$ 
which are orthogonal to $(e_0)^\mu_P$, we can further write 
\ba
\alpha_1^\mu = \frac{dx^\mu}{d\lambda}\Big|_{\lambda=0} 
=\:& x_F^i \left(e_i\right)_P^\mu.
\label{eq:def_xFi}
\ea
Since $(e_i)^\mu$ are parallel transported along the central geodesic,
they are uniquely defined along the geodesic once they are specified
at one point; i.e., they are unique up to three Euler angles.  
\refeq{def_xFi} defines the spatial Fermi coordinate $x_F^j$ (recall that
we have defined $\lambda$ through $x^\mu(\lambda=1) = Q$, so that
$\d_{ij} x_F^i x_F^j$ provides a measure for the spatial distance 
between points $Q$ and $P$).  

In order to obtain the metric in the Fermi coordinate up to 
including $\mathcal{O}(x_F^2)$, we also need quadratic ($\alpha_2^\mu$) 
and  cubic ($\alpha_3^\mu$) order coordinate transformation.
For that, we use the geodesic equation:
\be
\frac{d^2x^\mu}{d\lambda^2}
+
\Gamma^\mu_{\alpha\beta}
\frac{dx^\alpha}{d\lambda}
\frac{dx^\beta}{d\lambda}=0.
\ee
Rearranging the equation immediately yields
\ba
\left.\frac{d^2x^\mu}{d\lambda^2}\right|_{\lambda=0} 
=
2\alpha_2^\mu
=\:&
-\Gamma^\mu_{\alpha\beta}
\left.\frac{dx^\alpha}{d\lambda}
\frac{dx^\beta}{d\lambda}
\right|_{\lambda=0}
=
-\Gamma^\mu_{\alpha\beta}\Big|_P
(e_i)_P^\alpha (e_j)_P^\beta
x_F^ix_F^j \vs
\alpha_2^\mu
=\:&
-
\frac{1}{2}
\Gamma^\mu_{\alpha\beta}\Big|_P
(e_i)_P^\alpha (e_j)_P^\beta
x_F^ix_F^j.
\ea
Applying one more derivative with respect to $\lambda$ to the geodesic 
equation yields $\alpha_3^\mu$ through
\ba
\alpha_3^\mu
=\:&
\frac{1}{6}\left.\frac{d^3x^\mu}{d\lambda^3}\right|_{\lambda=0} 
=
-
\frac{1}{6}
\frac{d}{d\lambda}
\left(
\Gamma^\mu_{\alpha\beta}
\left.\frac{dx^\alpha}{d\lambda}
\frac{dx^\beta}{d\lambda}
\right)
\right|_{\lambda=0}
\vs
%=\:&
%-
%\frac{1}{6}
%\left(
%\Gamma^\mu_{\alpha\beta,\gamma}
%\left.\frac{dx^\alpha}{d\lambda}
%\frac{dx^\beta}{d\lambda}
%\frac{dx^\gamma}{d\lambda}
%\right|_{\lambda=0}
%+2
%\Gamma^\mu_{\alpha\beta}
%\left.
%\frac{d^2x^\alpha}{d\lambda^2}
%\frac{dx^\beta}{d\lambda}
%\right)
%\right|_{\lambda=0}
%\vs
%=\:&
%-
%\frac{1}{6}
%\left.
%\left(
%\Gamma^\mu_{\alpha\beta,\gamma}
%-2
%\Gamma^\mu_{\sigma\alpha}
%\Gamma^\sigma_{\beta\gamma}
%\right)
%\frac{dx^\alpha}{d\lambda}
%\frac{dx^\beta}{d\lambda}
%\frac{dx^\gamma}{d\lambda}
%\right|_{\lambda=0}
%\vs
=\:&
-
\frac{1}{6}
\left[
\Gamma^\mu_{\alpha\beta,\gamma}
-2
\Gamma^\mu_{\sigma\alpha}
\Gamma^\sigma_{\beta\gamma}
\right]_P
(e_i)_P^\alpha 
(e_j)_P^\beta
(e_k)_P^\gamma
x_F^ix_F^jx_F^k.
\ea
Combining all, we find the coordinate transformation
from FNC to general coordinates on a fixed $x^0_F$ hypersurface, 
up to third order in $x_F^i$:
\ba
x^\mu(x_F^i)\Big|_{t_P}
= x^\mu(P) +
\left(e_i\right)^\mu_P x_F^i
-
\frac{1}{2}
\Gamma^\mu_{\alpha\beta}\Big|_P
(e_i)^\alpha_P (e_j)^\beta_P
x_F^ix_F^j
-
\frac{1}{6}
\left[
\Gamma^\mu_{\alpha\beta,\gamma}
-2
\Gamma^\mu_{\sigma\alpha}
\Gamma^\sigma_{\beta\gamma}
\right]_P
(e_i)^\alpha_P 
(e_j)^\beta_P
(e_k)^\gamma_P
x_F^ix_F^jx_F^k.
\label{eq:FNCcoordTF}
\ea
Here we have made explicit that $(e_\nu)^\mu$ and $\G^\mu_{\alpha\beta}$
are always evaluated along the central geodesic at $P$.  
In \refapp{Fermimetric}, we use this to derive the quadratic order 
FNC metric $g_{\mu\nu}^F$ for a given global metric $g_{\mu\nu}$ 
[\refeq{Fermimetric}].  This result is also given in
Eq.~(66) of \cite{ManasseMisner} and Eq.~(13.73) in \cite{MTW}.  
In \refapp{Fermi_sc} we then apply this procedure to the perturbed FRW metric
in synchronous gauge [\refeq{metric}] to obtain the Fermi coordinates 
and corrections $\propto x_F^2$ to the metric.  
Readers familiar with Fermi coordinates and \refeq{Fermimetric} 
may want to skip to \refapp{Fermi_sc}.

%%%%%%%%%%%%%%%%%%%%%%%%%%%%%%%%%%%%%%%%%%%%%%%%%%%%%%%%%%%%%%%%%%%%%%%%%%%
\subsection{Metric in Fermi coordinate}
%%%%%%%%%%%%%%%%%%%%%%%%%%%%%%%%%%%%%%%%%%%%%%%%%%%%%%%%%%%%%%%%%%%%%%%%%
\label{app:Fermimetric}

In this section we derive the metric in the Fermi coordinate by using 
the coordinate transformation we have found in \refeq{FNCcoordTF}.  
The final result will be that derived by \cite{ManasseMisner},
\refeq{Fermimetric}.  
Under a coordinate transformation, the metric tensor transforms as
\be
g_{\mu\nu}^{F}(x_F)
=
\frac{\partial x^\alpha}{\partial x_F^\mu}
\frac{\partial x^\beta}{\partial x_F^\nu}
g_{\alpha\beta}(x).
\label{eq:gtransf}
\ee
Given the coordinate transform \refeq{FNCcoordTF}, we derive the 
partial derivatives to second order in $x_F$, yielding
\ba
\frac{\partial x^\mu}{\partial x^0_F}
=\:&
\frac{\partial x^\mu(P)}{\partial x^0_F}
+
\frac{\partial}{\partial x^0_F}(e_i)^\mu
x_F^i
-
\frac{1}{2}
\frac{\partial}{\partial x^0_F}
\left[
\Gamma^\mu_{\alpha\beta}\Big|_P
(e_i)^\alpha (e_j)^\beta
\right]
x_F^ix_F^j
%\vs
%=\:&
%(e_0)^\mu
%-
%\Gamma^\mu_{\alpha\beta}\Big|_P(e_i)^\alpha(e_0)^\beta x_F^i
%-
%\frac{1}{2}
%\left[
%\Gamma^\mu_{\alpha\beta,\nu}
%(e_0)^\nu (e_i)^\alpha (e_j)^\beta
%-
%2
%\Gamma^\mu_{\alpha\beta}
%\Gamma^\alpha_{\gamma\delta}
%(e_0)^\gamma (e_i)^\delta (e_j)^\beta
%\right]_P
%x_F^ix_F^j
\vs
=\:&
(e_0)^\mu
-
\Gamma^\mu_{\alpha\beta}\Big|_P(e_i)^\alpha(e_0)^\beta x_F^i
-
\frac{1}{2}
\left[
\Gamma^\mu_{\alpha\beta,\gamma}
-
2
\Gamma^\mu_{\sigma\beta}
\Gamma^\sigma_{\gamma\alpha}
\right]_P
(e_0)^\gamma (e_i)^\alpha (e_j)^\beta
x_F^ix_F^j
\label{eq:dxdtF}
\\
\frac{\partial x^\mu}{\partial x_F^l}
=\:&
(e_l)^\mu
-
\Gamma^\mu_{\alpha\beta}\Big|_P
(e_i)^\alpha (e_l)^\beta
x_F^i
-
\frac{1}{6}
\left[
\Gamma^\mu_{\alpha\beta,\gamma}
-2
\Gamma^\mu_{\sigma\gamma}
\Gamma^\sigma_{\alpha\beta}
\right]_P
\left[
(e_i)^\alpha 
(e_j)^\beta
(e_l)^\gamma
x_F^ix_F^j
+
2
(e_l)^\alpha 
(e_j)^\beta
(e_k)^\gamma
x_F^jx_F^k
\right]
\vs
=\:&
(e_l)^\mu
-
\Gamma^\mu_{\alpha\beta}\Big|_P
(e_i)^\alpha (e_l)^\beta
x_F^i
-
\frac{1}{6}
\left[
\Gamma^\mu_{\alpha\beta,\gamma}
+
2\Gamma^\mu_{\gamma\a,\b}
-2
\Gamma^\mu_{\sigma\gamma}
\Gamma^\sigma_{\alpha\beta}
-4
\Gamma^\mu_{\sigma\b}
\Gamma^\sigma_{\gamma\a}
\right]_P
(e_i)^\alpha 
(e_j)^\beta
(e_l)^\gamma
x_F^ix_F^j,
\label{eq:dxdxF}
\ea
where all unit vectors are evaluated at $P$.  
Note that $\partial x^\mu(P)/\partial x^0_F = (e_0)^\mu_P$, since
by definition $(e_0)^\mu$ is the tangent vector to the central geodesic 
at $P$.  Further, we have used that $(e_i)^\mu_P$ and 
$\Gamma^\mu_{\alpha\beta}\Big|_P$ only depend on $x^0_F$, and that 
by construction, the basis vectors
$(e_i)^\mu$ are parallel transported along the central geodesic.  This implies
\ba
0 = \frac{D}{Dx^0_F}(e_i)^\a
=
(e_i)_{;\mu}^\a
(e_0)^\mu
=\:&
\left[
\frac{\partial}{\partial x^\mu} (e_i)^\alpha 
+
\Gamma^{\alpha}_{\beta\mu} (e_i)^\beta
\right]
(e_{0})^\mu
=
\frac{\partial}{\partial x^0_F} (e_i)^\alpha 
+
\Gamma^{\alpha}_{\beta\mu} (e_i)^\beta (e_0)^\mu \vs
\Rightarrow \frac{\partial}{\partial x^0_F} (e_i)^\alpha 
=\:& - \Gamma^{\alpha}_{\beta\mu} (e_i)^\beta (e_0)^\mu.
\ea
In \refeq{dxdtF} we have also used
\be
\frac{\partial}{\partial x^0_F}
\Gamma^\mu_{\a\b}\Big|_P
=
\Gamma^\mu_{\a\b,\gamma}\Big|_P \left(e_0\right)_P^\gamma.
\ee
Finally, we need to take into account that the metric on the right-hand
side of \refeq{gtransf} is evaluated at a point $Q$ (specified by $x_F^i$)
away from the central geodesic.  We perform a Taylor expansion of
$g_{\mu\nu}$ around $P$,
\ba
g_{\alpha\beta}(Q)
=\:&
g_{\alpha\beta}\Big|_P
+
g_{\alpha\beta,\mu}\Big|_P \delta x^\mu
+
\frac12
g_{\alpha\beta,\mu\nu}\Big|_P \delta x^\mu \delta x^\nu
+
\mathcal{O}(\delta x^3)
\ea
where, from \refeq{FNCcoordTF},
%\ba
%\delta x^0
%=\:&
%-
%\frac12\Gamma^0_{\alpha\beta}\Big|_P(e_i)_P^\alpha(e_j)_P^\beta x_F^i x_F^j
%\\
%\delta x^i
%=\:&
%(e_j)^ix_F^j
%-
%\frac12\Gamma^i_{\alpha\beta}\Big|_P(e_j)_P^\alpha(e_k)_P^\beta x_F^j x_F^k.
%\ea
\be
\delta x^\mu
=
\left(e_i\right)^\mu_P x_F^i
-
\frac12 \G^\mu_{\alpha_\beta}
\left(e_i\right)^\a_P 
\left(e_j\right)^\b_P 
x_F^i
x_F^j
\ee
That is, up to second order in $x_F^i$, the metric at $Q$ is given by
\ba
g_{\alpha\beta}(Q)
=\:&
g_{\alpha\beta}\Big|_P
+
g_{\alpha\beta,\mu}\Big|_P
(e_i)_P^\mu x_F^i
+
\frac12
\left[
g_{\alpha\beta,\mu\nu}
-
g_{\alpha\beta,\sigma}
\Gamma^\sigma_{\mu\nu}
\right]_P
(e_i)_P^\mu (e_j)_P^\nu x_F^i x_F^j.
\label{eq:gPgQ}
\ea
From now on, every instance of $g_{\mu\nu}$, $\G^\sigma_{\alpha\beta}$, and
$(e_\nu)^\mu$ 
will be evaluated at $P$, and we will omit $P$ for brevity hereafter.  
Inserting \refeq{dxdtF}, \refeq{dxdxF} and \refeq{gPgQ} into 
\refeq{gtransf}, and expanding to second order in $x_F^i$ yields the 
desired metric in FNC.  At linear order, \refeq{gtransf} becomes
\ba
g_{\mu\nu}^{F}
=\:&
\biggl[
\left(e_\mu\right)^\a 
- \Gamma^\a_{\rho\sigma}\left(e_i\right)^\rho\left(e_\mu\right)^\sigma x_F^i
\biggl]
\biggl[
\left(e_\nu\right)^\b 
- \Gamma^\b_{\rho\sigma}\left(e_j\right)^\rho\left(e_\nu\right)^\sigma x_F^j
\biggl]
\left[
g_{\alpha\beta}
+
g_{\alpha\beta,\k}(e_k)^\k x_F^k
\right]
\vs
=\:&
\left(e_\mu\right)^\a
\left(e_\nu\right)^\b
g_{\alpha\beta}
+
\left(
g_{\alpha\beta,\rho}
- 
g_{\sigma\beta}
\Gamma^\sigma_{\rho\a}
-
g_{\alpha\sigma}
\Gamma^\sigma_{\rho\b}
\right)
\left(e_\mu\right)^\a
\left(e_\nu\right)^\b
\left(e_i\right)^\rho
x_F^i
\vs
=\:&
\left(e_\mu\right)^\a
\left(e_\nu\right)^\b
g_{\alpha\beta}
+
g_{\alpha\beta;\rho}
\left(e_\mu\right)^\a
\left(e_\nu\right)^\b
\left(e_i\right)^\rho
x_F^i
=
\eta_{\mu\nu},
\ea
where the last equality follows 
from the definition of the orthonormal tetrad at $P$,
\be
g_{\mu\nu} (e_\a)^\mu (e_\b)^\nu = \eta_{\a\b},
\ee
and the Levi-Civita connection, $g_{\mu\nu;\rho}=0$.  

Next, we calculate the quadratic correction to the metric in FNC, 
\ba
\label{eq:2ndFermi1}
\delta g_{00}^{F}
=\:&
\biggl[
\frac12
g_{\mu\nu,\a\b}
-
\frac12
g_{\mu\nu,\sigma}
\Gamma^\sigma_{\a\b}
-
2g_{\mu\gamma,\a}
\Gamma^\gamma_{\b\nu}
-
g_{\mu\gamma}
\Gamma^\gamma_{\alpha\beta,\nu}
+
2
g_{\mu\gamma}
\Gamma^\gamma_{\sigma\beta}
\Gamma^\sigma_{\nu\alpha}
+
g_{\gamma\sigma}\G^\gamma_{\a\mu}\G^\sigma_{\b\nu}
\biggl]
(e_0)^\mu
(e_0)^\nu
(e_l)^\a (e_m)^\b x_F^l x_F^m
\\
\delta g_{0i}^{F}
=\:&
\biggl[
\frac12g_{\mu\nu,\a\b}
-\frac12g_{\mu\nu,\sigma}\G^\sigma_{\a\b}
-
g_{\mu\gamma,\a} \G^{\gamma}_{\b\nu}
-
g_{\nu\gamma,\a} \G^{\gamma}_{\b\mu}
-\frac12g_{\nu\gamma}\G^{\gamma}_{\a\b,\mu}
+g_{\nu\gamma}\G^\gamma_{\sigma\a}\G^\sigma_{\mu\b}
+
g_{\gamma\sigma}\G^\gamma_{\a\mu}\G^\sigma_{\b\nu}
\vs
&-\frac16 g_{\mu\lambda}
\left(
\G^\lambda_{\a\b,\nu}
+2
\G^\lambda_{\nu\a,\b}
-2
\G^{\lambda}_{\sigma\nu} \G^\sigma_{\a\b}
-4
\G^{\lambda}_{\sigma\b} \G^\sigma_{\a\nu}
\right)
\biggl]
(e_0)^\mu
(e_i)^\nu
(e_l)^\a (e_m)^\b x_F^l x_F^m
\label{eq:2ndFermi2}
\\
\delta g_{ij}^{F}
=\:&
\biggl[
\frac12 g_{\mu\nu,\a\b}
-
\frac12
g_{\mu\nu,\sigma}\G^\sigma_{\a\b}
-
2g_{\mu\gamma,\a}\G^\gamma_{\nu\b}
+
g_{\gamma\sigma}\G^\gamma_{\a\mu}\G^\sigma_{\b\nu}
\vs
&-\frac13 g_{\mu\lambda}
\left(
\G^\lambda_{\a\b,\nu}
+2
\G^\lambda_{\nu\a,\b}
-2
\G^{\lambda}_{\sigma\nu} \G^\sigma_{\a\b}
-4
\G^{\lambda}_{\sigma\b} \G^\sigma_{\a\nu}
\right)
\biggl]
(e_i)^\mu
(e_j)^\nu
(e_l)^\a (e_m)^\b x_F^l x_F^m.
\label{eq:2ndFermi3}
\ea
Finally, using
\be
0 = g_{\mu\nu;\a}
=
g_{\mu\nu,\a}
-
g_{\sigma\nu}\G^\sigma_{\mu\a}
-
g_{\mu\sigma}\G^\sigma_{\nu\a},
\label{eq:D1g}
\ee
we have
\be
g_{\mu\nu,\a\b}
=
g_{\mu\sigma}\G^\sigma_{\nu\a,\b}
+
g_{\sigma\nu}\G^\sigma_{\mu\a,\b}
+
g_{\mu\g}\G^\g_{\sigma\b}\G^\sigma_{\nu\a}
+
g_{\g\nu}\G^{\g}_{\sigma\b}\G^\sigma_{\mu\a}
+
g_{\g\sigma}
\left(
\G^{\g}_{\mu\b}\G^{\sigma}_{\nu\a}
+
\G^{\g}_{\nu\b}\G^{\sigma}_{\mu\a}
\right).
\label{eq:D2g}
\ee
By using \refeqs{D1g}{D2g}, 
we write the partial derivatives of the metric as a function of 
the metric itself and the Christoffel symbols, to obtain
the final expression for the FNC metric at quadratic order:
\ba
\delta g_{00}^{F}
=\:&
%R_{\mu\a\b\nu}
%(e_0)^\mu
%(e_0)^\nu
%(e_l)^\a (e_m)^\b 
%x_F^l x_F^m
%=
-R^{F}_{0l0m}
x_F^l x_F^m
\label{eq:Fermimetric}\\
\delta g_{0i}^{F}
=\:&
%\left[
%\frac12 R_{\nu\a\b\mu}
%+
%\frac16 R_{\mu\a\b\nu}
%\right]
%(e_0)^\mu
%(e_i)^\nu
%(e_l)^\a (e_m)^\b 
%x_F^l x_F^m
%=
-\frac23 R^{F}_{0lim}
x_F^l x_F^m
\vs
\delta g_{ij}^{F}
=\:&
%\frac13 R_{\mu\a\b\nu}
%(e_i)^\mu
%(e_j)^\nu
%(e_l)^\a (e_m)^\b 
%x_F^l x_F^m
%=
-\frac13 R^{F}_{iljm}
x_F^l x_F^m.\nonumber
\ea
Here, we have defined $R^F_{\alpha\beta\gamma\delta}$ as the
Riemann tensor in FNC,
\be
R^{F}_{\alpha\beta\gamma\delta} = (e_\alpha)^\mu (e_\beta)^\nu
(e_\gamma)^\kappa (e_\delta)^\lambda\,R_{\mu\nu\kappa\lambda},
\label{eq:transRiemann}
\ee
where the Riemann tensor is defined following the convention of
\cite{MTW}
\be
R^\mu_{\;\;\alpha\beta\gamma} = \G^\mu_{\alpha\gamma,\beta} 
- \G^\mu_{\alpha\beta,\gamma} 
+ 
\G^\mu_{\sigma\beta}\G^\sigma_{\alpha\gamma}
- 
\G^\mu_{\sigma\gamma}\G^\sigma_{\alpha\beta}.
\label{eq:Riemann}
\ee

%%%%%%%%%%%%%%%%%%%%%%%%%%%%%%%%%%%%%%%%%%%%%%%%%%%%%%%%%%%%%%%%%%%%%%%%%%%
\subsection{Application to synchronous gauge metric}
\label{app:Fermi_sc}
%%%%%%%%%%%%%%%%%%%%%%%%%%%%%%%%%%%%%%%%%%%%%%%%%%%%%%%%%%%%%%%%%%%%%%%%%

We write the perturbed FRW metric \refeq{metric} in terms of proper 
time $t$ instead of conformal time $\eta$,
\be
ds^2 = -dt^2 + a^2(t)[\d_{ij} + h_{ij}] dx^i dx^j.
\label{eq:metricG}
\ee
For this metric, coordinate time coincides with proper time for
a comoving observer ($x^i =$~const).  
Thus, without loss of generality we choose
$\{(t,0,0,0)\}_t$ as central geodesic.  Further, since the FNC time 
coordinate $t_F$ is given by the proper time along the central geodesic,
we have $t_F = t$ along the central geodesic (while $t_F\neq t$ 
for $x_F^i \neq 0$).  Correspondingly, the unit time vector is given by
$(e_0)^\mu = (1,0,0,0)$, and orthogonal spatial basis vectors
are given by (see also \cite{gaugePk})
\ba
(e_k)^\mu =\:& \left(0, \frac1a 
\left[\d_{ik} - \frac12 h_{ik}\right]\right).
\ea
The inverse metric is 
\be
g^{00} = -1,~
g^{ij} = \frac{1}{a^2}
\left(\delta_{ij}-h_{ij}\right).
\ee
This leads to the following Christoffel symbols,
\ba
\G^0_{\ 00} =\:& \G^0_{\ 0i} = \G^i_{00} = 0
\\
\G^0_{\ ij} 
=\:& a^2 H \d_{ij} + a^2 H h_{ij} + \frac{a^2}{2}\dot{h}_{ij}
\\
\G^i_{\ 0j} 
=\:& H \d_{ij} +\frac{1}{2}\dot{h}_{ij}
\\
\G^i_{\ jk} 
=\:& \frac12
\left(
h_{ji,k} + h_{ki,j} - h_{jk,i}
\right),
\ea
where here and throughout, dots denote derivatives with respect to $t$.  
The Riemann tensor is given by
\ba
R^i_{\ 00m}
=\:&
\left(\dot{H} + H^2\right) \delta_{im}
+\frac{1}{2}\ddot{h}_{i m}
+
H\dot{h}_{im}
\\
R^{n}_{\ 0im}
=\:&
\frac{1}{2}
\left(
\dot{h}_{nm,i}
-
\dot{h}_{ni,m}
\right)
\\
R^n_{\ ijm}
=\:&
a^2H^2\left[\d_{nj}\d_{im}-\d_{nm}\d_{ij}\right]
+
\frac12
\left(
h_{mn,ij} + h_{ij,nm}
- h_{im,nj} -h_{jn,im} 
\right)
\vs
&+
a^2H^2\left(h_{im}\d_{nj}-h_{ij}\d_{nm}\right)
+
\frac{a^2H}{2}
\left(
\dot{h}_{nj}\d_{im}
+
\dot{h}_{im}\d_{nj}
-
\dot{h}_{ij}\d_{nm}
-
\dot{h}_{nm}\d_{ij}
\right).
\ea 
We then have
\ba
R_{i00m}
=\:&
g_{ij}R^{j}_{\ 00m}
=
a^2\left(\dot{H}+H^2\right)\delta_{im}
+a^2
\left[
\frac{1}{2}\ddot{h}_{im}
+
H\dot{h}_{im}
+
\left(\dot{H}+H^2\right)
h_{im}
\right]
\\
R_{l0im}
=\:&
\frac{a^2}{2}
\left(
\dot{h}_{lm,i}
-
\dot{h}_{li,m}
\right)
\\
R_{lijm}
=\:&
a^4H^2\left[\d_{lj}\d_{im}-\d_{lm}\d_{ij}\right]
+\frac{a^2}{2}
\left(
h_{ml,ij} + h_{ij,lm}
- h_{im,lj} -h_{jl,im} 
\right)
\vs
&+
a^4H^2\left(
h_{im}\d_{lj}
+h_{lj}\d_{im}
-h_{ij}\d_{lm}
-h_{lm}\d_{ij}
\right)
+
\frac{a^4H}{2}
\left(
\dot{h}_{lj}\d_{im}
+
\dot{h}_{im}\d_{lj}
-
\dot{h}_{ij}\d_{lm}
-
\dot{h}_{lm}\d_{ij}
\right).
\ea
Finally, the Riemann tensor in terms of FNC is given by
\ba
R^{F}_{l00m} 
=\:& 
\left(e_l\right)^\mu
\left(e_0\right)^\nu
\left(e_0\right)^\k
\left(e_m\right)^\lambda R_{\mu\nu\k\lambda}
=
\left(\dot{H}+H^2\right)\delta_{lm}
+
\left[
\frac{1}{2}\ddot{h}_{lm}
+
H\dot{h}_{lm}
\right]
\\
R^{F}_{l0im} 
=\:& 
\frac{1}{2a}\left(
\dot{h}_{lm,i} - \dot{h}_{li,m}
\right)
\\
R^{F}_{lijm} =\:& 
H^2\left[\d_{lj}\d_{im}-\d_{lm}\d_{ij}\right]
+\frac{1}{2a^2}
\left(
h_{ml,ij} + h_{ij,lm}
- h_{im,lj} -h_{jl,im} 
\right)
+
\frac{H}{2}
\left(
\dot{h}_{lj}\d_{im}
+
\dot{h}_{im}\d_{lj}
-
\dot{h}_{ij}\d_{lm}
-
\dot{h}_{lm}\d_{ij}
\right).
\ea
Combining all with \refeq{Fermimetric}, we find that the metric in FNC is 
\ba
g_{00}^{F} 
=\:&
-1
+
\left(\dot{H}+H^2\right)
r_F^2
+
\left[
\frac{1}{2}\ddot{h}_{lm}
+
H\dot{h}_{lm}
\right]x^l_Fx^m_F.
\vs
g_{0i}^{F}
=\:&
\frac{1}{3}\left(
\nabla_i\dot{h}_{lm} - \nabla_m\dot{h}_{li}
\right) x_F^l x_F^m
\vs
g_{ij}^{F}
=\:& \:
\delta_{ij}
+
\frac{
H^2}{3}
\left[
x_F^ix_F^j
-r_F^2\d_{ij}
\right]
+
\frac{1}{6}
\left(
\nabla_i\nabla_j h_{ml} 
+ 
\nabla_l\nabla_m h_{ij}
- 
\nabla_l\nabla_j
h_{im} 
-
\nabla_i\nabla_m
h_{jl} 
\right) x_F^lx_F^m
\vs
&\: +
\frac{H}{6}
\left(
\dot{h}_{lj}x_F^lx_F^i
+
\dot{h}_{im}x_F^mx_F^j
-
\dot{h}_{ij}r_F^2
-
\dot{h}_{lm}x_F^lx_F^m\d_{ij}
\right).
\label{eq:Fermimetric_F}
\ea
Here, we define $r_F^2 = \delta_{ij} x_F^ix_F^j$ and 
denote the partial derivative with respect to the FNC by
$\nabla_i \equiv \partial/\partial x_F^i$.  Note that
in \refeq{Fermimetric_F}, the derivative terms are already order
$x_F^2$, hence we can use $\nabla_i = (1/a)\partial/\partial x^i$ here.  
We reiterate that \refeq{Fermimetric_F} is valid for any spatial
metric perturbation $h_{ij}$, and thus also encompasses scalar 
cosmological perturbations written in synchronous-comoving gauge.  

It is also useful to have an explicit expression for the transformation
from global coordinates $x^\mu$ to Fermi coordinates $x_F^\nu$. 
Evaluating \refeq{FNCcoordTF} for the metric \refeq{metricG} yields
\ba
x^i =\:&
\left(\delta_{ij} -\frac12 h_{ij}\right) \frac1a
x_F^j
-
\frac12 \G^i_{jk} \frac1{a^2}
x_F^j
x_F^k + \O(x_F^3)
\label{eq:xFtox} \\
\frac1a x_F^i =\:& 
\left(\delta_{ij} + \frac12 h_{ij}\right) 
x^j
+
\frac12 \G^i_{jk}
x^j
x^k + \O(x^3)
\label{eq:xtoxF}.
\ea
This result is used in \S~VIIB of \cite{stdruler}.  

Finally, given the FNC metric \refeq{Fermimetric_F}, we can derive 
the motion of non-relativistic bodies (of momentum $p^i$ and mass $m$), 
which is governed by
\ba
\frac1m\frac{dp^i}{dt} 
=\:&
\frac12 \,  g^{F}_{00,i}
= (H^2+\dot H) x_F^i - \nabla_i \Psi^F
\label{eq:eomFermi}\\
\Psi^F =\:& -\frac12
\left(\frac12 \ddot{h}_{ij} + H\dot{h}_{ij}\right)
x_F^ix_F^j.
\ea
This is the usual quasi-Newtonian equation of motion of a particle
in an expanding Universe with peculiar potential $\Psi^F$.  
The effective potential induces a tidal tensor given by
\ba
t_{ij}
\equiv& 
\left(\partial_i\partial_j - \frac13 \delta_{ij}\nabla^2\right) 
\Psi^F
\vs
=\:&
-\frac12
\left(\frac12 \ddot{h}_{lm} + H\dot{h}_{lm}\right)
\left(\partial_i\partial_j - \frac13 \delta_{ij}\nabla^2\right) 
x_F^lx_F^m
\vs
=\:&
-\left[
\left(\frac12 \ddot{h}_{lm} + H\dot{h}_{lm}\right)
-\frac13\delta_{ij}
{\rm Tr}\left(\frac12 \ddot{h}_{lm} + H\dot{h}_{lm}\right)
\right].
\ea
If $h_{ij}$ is traceless, the last term vanishes and we obtain
\be
t_{ij} \stackrel{\rm traceless}{=}
-\left(\frac12 \ddot{h}_{lm} + H\dot{h}_{lm}\right).
\ee

%%%%%%%%%%%%%%%%%%%%%%%%%%%%%%%%%%%%%%%%%%%%%%%%%%%%%%%%%%%%%%%%%%%%%%%%%%%
\section{Derivation of shear}
\label{app:shear}

% % % % % % % % % % % % % % % % % % % % % % % % % % % % % % % % % % % % %
\subsection{Shear statistics from tensor modes}

We express the tensor metric perturbation $h_{ij}(\tilde\vx,\eta)$ as
\ba
h_{ij}(\tilde\vx,\eta) =\:& \int \frac{d^3k}{(2\pi)^3} \left[
e^+_{ij}(\hat\vk) h^+(\vk,\eta) + e^\times_{ij}(\hat\vk)
h^\times(\vk, \eta)
\right] e^{i\vk\cdot\vnhat\,\chit} \vs
=\:& \int \frac{d^3k}{(2\pi)^3} \sum_{p=-1,1}
e^p_{ij}(\hat\vk) h_p(\vk,\eta) \:e^{i\vk\cdot\vnhat\,\chit},
\ea
where we have defined the helicity$\pm 2$ polarization tensors and
Fourier amplitudes through
\ba
e^{\pm1}_{ij} \equiv\:& e^+_{ij} \pm i e^\times_{ij} \vs
h_{\pm1} \equiv\:& \frac12 (h_+ \mp i h_\times).
\ea
In standard spherical coordinates, we have
\ba
m^i_\pm =\:& 
\frac{1}{\sqrt{2}}
\left(
\hat{e}_\theta
\mp i \hat{e}_\phi
\right)
=
\frac{1}{\sqrt{2}}\left(\begin{array}{c}
\cos\theta \cos\phi \pm i \sin\phi\\
\cos\theta \sin\phi \mp i \cos\phi\\
-\sin\theta
\end{array}\right),
\ea
where $\hat{e}_\theta$ and $\hat{e}_\phi$ are, respectively, the unit vectors
of the polar and azimuthal angles.  
In order to make progress, we begin by evaluating the contribution of
a single plane wave, assuming that $\vk = k \hat{\v{z}}$.  We have
\ba
e^p_{\pm}(\hat\vk,\vnhat) \equiv e^p_{ij}(\hat\vk) m_\mp^i(\vnhat) m_\mp^j(\vnhat) =\:&  
\frac12 (1 \mp p \mu)^2 e^{i2p\phi}
\vs
e^p_\parallel(\hat\vk) \equiv e^p_{ij} \nhat^i \nhat^j  =\:& 
(1-\mu^2) e^{i2p\phi}
\vs
e^p_{ij}(\hat\vk) m_\mp^i(\vnhat) \nhat^j =\:& 
\frac{\sqrt{1-\mu^2}}{\sqrt2} (\mu\mp p) e^{i2p\phi},
\label{eq:econtr}
\ea
where $p=\pm1$ and $\mu=\cos\theta$.  
We will also use the notation $k_\pm = m_{\mp}^i k_i = -\sin\theta\,k/\sqrt{2}$. 
Using \refeq{shear11} and \refeq{shearIA}, we then have for the contribution to the shear
\ba
({}_{\pm2}\g)(\vk,\vnhat) = \sum_{p=-1,1}\Bigg\{ &
- \frac12 \left[ h_p(\vk,\eta_0) + h_p(\vk,\tilde\eta) e^{i\vk\cdot\vnhat\,\chit}\right] e^p_\pm 
 + \frac13 C_1\rhocr H_0^{-2} \tilde a^{-2}
\left(h''_p(\vk,\tilde\eta) + \tilde a\tilde H h'_p(\vk,\tilde\eta)\right)
e^{i\vk\cdot\vnhat\,\chit} e^p_\pm
\vs
 & - \int_0^{\chit} d\chi
\bigg[
\frac{\chit-\chi}{2}\frac{\chi}{\chit}
(- k_\pm^2) e^p_\parallel
+ \left(1-2\frac{\chi}{\chit}\right) 
i k_\pm m_\mp^k \nhat^l e^p_{kl} 
- \frac1{\chit} e^p_\pm
\bigg] h_p(\vk,\eta_0-\chi) e^{i\vk\cdot\vnhat\,\chi}\Bigg\} \vs
%%%
=  \sum_{p=-1,1}\Bigg\{ &
- \frac12 \left[ h_p(\vk,\eta_0) + 
\left(1 - \frac23 C_1\rhocr H_0^{-2}\tilde a^{-2} \left\{\partial_{\tilde\eta}^2 + \tilde a\tilde H \partial_{\tilde\eta}\right\}\right)
h_p(\vk,\tilde\eta) e^{i\vk\cdot\vnhat\,\chit}\right] \frac12 (1\mp p\mu)^2 e^{i2p\phi} \vs
 & + \int_0^{\chit} d\chi
\bigg[
\frac{\chit-\chi}{4}\frac{\chi}{\chit}
k^2 (1-\mu^2)^2
+ \left(1-2\frac{\chi}{\chit}\right) 
i \frac{k}2 (1-\mu^2) (\mu\mp p)
+ \frac1{2\chit}(1\mp p\mu)^2 
\bigg] \vs
& \hspace*{1cm} \times e^{i2p\phi} h_p(\vk,\eta_0-\chi) e^{i\vk\cdot\vnhat\,\chi}\Bigg\}. \vs
\ea
Here, $\tilde\eta = \eta_0-\chit$ is the conformal time at emission inferred
from the observed redshift, and all tilded quantities are evaluated at the
source redshift.  The next step will be to derive the
spherical harmonic coefficients of the shear.  Clearly, all factors involve
$e^{\pm i2\phi}$, so that only spherical harmonic coefficients with $m=\pm2$ 
will be non-zero (this is of course a consequence of the choice $\vk=k\hat{\v{z}}$).  

% % % % % % % % % % % % % % % % % % % % % % % % % % % % % % % % % % % %
\subsection{Spin-raising and lowering}
\label{app:shear2}

Let us consider the case of ${}_{2}\g$, and restrict to one circular 
polarization $p=+1$ first:
\ba
{}_2\g(\vk,\vnhat,+1) =\:&
- \frac14 \left[ h_1(\vk,\eta_0) e^{i x\mu}\Big|_{x=0} 
+ \left(1 - \frac23 C_1\rhocr H_0^{-2}\tilde a^{-2} \left\{\partial_{\tilde\eta}^2 + \tilde a\tilde H \partial_{\tilde\eta}\right\}\right)
h_1(\vk,\tilde\eta) e^{i \tilde x \mu}\right] (1 - \mu)^2 e^{i2\phi} \vs
 & + \int_0^{\chit} d\chi
\bigg[
\frac{\chit-\chi}{4}\frac{\chi}{\chit}
k^2 (1-\mu^2)^2
+ \left(1-2\frac{\chi}{\chit}\right) 
\frac{ix}{2\chi} (1-\mu^2) (\mu-1)
+ \frac1{2\chit}(1 - \mu)^2 
\bigg] e^{i2\phi} h_1(\vk,\eta_0-\chi) e^{i x \mu},
\label{eq:gp1}
\ea
where we have defined $x = k\chi$, $\tilde x = k\chit$.  
Since this is a spin$+2$ quantity, we apply the spin-lowering operator
twice to obtain a scalar.  For this, we use that for functions that
satisfy $\partial_\phi\, {}_s f = i m\, {}_s f$ (see App.~A of \cite{stdruler}
and \cite{ZalSel97} for details),
\ba
\bar\Del^2\: {}_2 f(\mu,\phi) =\:& \left(-\frac{\partial}{\partial\mu} + \frac{m}{1-\mu^2}\right)^2 \left[ (1-\mu^2)\:{}_2 f(\mu,\phi) \right]
\vs
\Del^2\: {}_{-2} f(\mu,\phi) =\:& \left(-\frac{\partial}{\partial\mu} - \frac{m}{1-\mu^2}\right)^2 \left[ (1-\mu^2)\:{}_{-2} f(\mu,\phi) \right].
\label{eq:Del}
\ea
With $m=2$, this yields
\ba
\bar\Del^2\,{}_2\g(\vk,\vnhat,+1) =\:&
- \frac14 \left(-\frac{\partial}{\partial\mu} + \frac{2}{1-\mu^2}\right)^2 
\bigg\{ \vs
& \qquad\qquad (1-\mu^2) (1 - \mu)^2    
\left[ h_1(\vk,\eta_0) e^{i x \mu}\Big|_{x=0}\!\!\!
+ \left(1 - \frac23 C_1\rhocr H_0^{-2}\tilde a^{-2} \left\{\partial_{\tilde\eta}^2 + \tilde a\tilde H \partial_{\tilde\eta}\right\}\right)
h_1(\vk,\tilde\eta) e^{i \tilde x \mu }\right]\bigg\} 
e^{i2\phi} \vs
 & + \int_0^{\chit} d\chi
\left(-\frac{\partial}{\partial\mu} + \frac{2}{1-\mu^2}\right)^2 (1-\mu^2)
\bigg[
\frac14 \frac{\chit-\chi}{\chi \chit}
x^2 (1-\mu^2)^2
+ \frac12 \frac{\chit-2\chi}{\chi \chit}
i x (1-\mu^2) (\mu-1)
+ \frac1{2\chit}(1 - \mu)^2 
\bigg] \vs
& \hspace{5.7cm} \times e^{i2\phi} h_1(\vk,\eta_0-\chi) e^{i x \mu} \vs
%%%
=\:& - \frac14 \left(-\frac{\partial}{\partial\mu} + \frac{2}{1-\mu^2}\right)^2 
\bigg\{ \vs
& \qquad\qquad (1-\mu^2) (1 - \mu)^2    
\left[ h_1(\vk,\eta_0) e^{i x \mu}\Big|_{x=0}\!\!\!
+ \left(1 - \frac23 C_1\rhocr H_0^{-2}\tilde a^{-2} \left\{\partial_{\tilde\eta}^2 + \tilde a\tilde H \partial_{\tilde\eta}\right\}\right)
h_1(\vk,\tilde\eta) e^{i \tilde x \mu }\right]\bigg\} 
e^{i2\phi} \vs
 & + \int_0^{\chit} \frac{d\chi}{\chi}
\left(-\frac{\partial}{\partial\mu} + \frac{2}{1-\mu^2}\right)^2 (1-\mu^2)
\bigg[
\left( \frac14 x^2 (1-\mu^2)^2 + \frac12 i x (1-\mu^2) (\mu-1)
\right) \vs
& \hspace*{6cm}
+ \frac{\chi}{\chit} \left(
- \frac14 x^2 (1-\mu^2)^2 - i x (1-\mu^2) (\mu-1) + \frac12 (1 - \mu)^2 
\right)
\bigg] \vs
& \hspace{5.5cm} \times e^{i2\phi} h_1(\vk,\eta_0-\chi) e^{i x \mu}\vs
%%%
=\:&
\Bigg\{ - \frac14 
\left[ h_1(\vk,\eta_0) \left(\hat Q_1(x) e^{i x \mu}\right)\Big|_{x=0}
+ \left(1 - \frac23 C_1\rhocr H_0^{-2}\tilde a^{-2} \left\{\partial_{\tilde\eta}^2 + \tilde a\tilde H \partial_{\tilde\eta}\right\}\right)
h_1(\vk,\tilde\eta) \hat Q_1(\tilde x) e^{i \tilde x \mu }\right]
\vs
 & \quad+ \int_0^{\chit} \frac{d\chi}{\chi}
\bigg[ \hat Q_2(x) + \frac{\chi}{\chit} \hat Q_3(x)
\bigg] h_1(\vk,\eta_0-\chi) e^{i x \mu}
\Bigg\} (1-\mu^2) e^{i2\phi},
\ea
where we have turned powers of $\mu$ into powers of $-i \partial_x$
and defined derivative operators $\hat Q_i(x)$ as  
\ba
\hat Q_1(x) =\:& 12 - x^2 + 8x \partial_x + x^2 \partial_x^2
- i\left( 8x + 2 x^2 \partial_x \right)
\vs
\hat Q_2(x) =\:& -\frac14\left[14x^2+x^4 + (40x+14 x^3) \partial_x
+ (50x^2+2x^4)\partial_x^2 + 14 x^3 \partial_x^3 + x^4 \partial_x^4 \right]
- \frac12 i \left[4x+x^3 + 6x^2 \partial_x + x^3\partial_x^2\right] \vs
\hat Q_3(x) =\:& \frac14\left[
24+24 x^2 + x^4 + (96x+16x^3) \partial_x + (72x^2+2x^4)\partial_x^2
+ 16x^3 \partial_x^3 + x^4 \partial_x^4 \right].
\ea
Although these operators are complicated, they will facilitate the
connection with the convergence and rotation below.  
Note that, as expected, the real parts of $\hat Q_i(x)$ only involve even powers
of $x$ (counting derivatives as well), 
while the imaginary parts involve odd powers only (where $\Im \hat Q_3 =0$).  
We now turn to $p=-1$:
\ba
{}_2\g(\vk,\vnhat,-1) =\:&
- \frac14 \left[ h_{-1}(\vk,\eta_0) e^{i x\mu}\Big|_{x=0} 
+ \left(1 - \frac23 C_1\rhocr H_0^{-2}\tilde a^{-2} \left\{\partial_{\tilde\eta}^2 + \tilde a\tilde H \partial_{\tilde\eta}\right\}\right)
h_{-1}(\vk,\tilde\eta) e^{i \tilde x \mu}\right] (1 + \mu)^2 e^{-i2\phi} \vs
 & + \int_0^{\chit} d\chi
\bigg[
\frac{\chit-\chi}{4}\frac{\chi}{\chit}
k^2 (1-\mu^2)^2
+ \left(1-2\frac{\chi}{\chit}\right) 
\frac{-ix}{2\chi}
(1-\mu^2) (-\mu - 1)
+ \frac1{2\chit}(1 + \mu)^2 
\bigg] e^{-i2\phi} h_{-1}(\vk,\eta_0-\chi) e^{i x \mu}.
\ea
The appropriate spin-lowering operator is now for $m=-2$,
i.e. 
\be
(-\partial_\mu - 2/(1-\mu^2))^2 = (\partial_\mu + 2/(1-\mu^2))^2
= (-\partial_{-\mu} + 2/(1-\mu^2))^2.
\ee
Thus, ${}_2\g(\vk,\vnhat,-1)$ is equal to ${}_2\g(\vk,\vnhat,+1)$
[\refeq{gp1}] when changing $\mu\to-\mu$, $x\to-x$ in addition to
$h_1 \to h_{-1},\:\phi\to-\phi$.  
Since $\hat Q_i(-x) = \hat Q_i^*(x)$, we obtain
\ba 
\bar\Del^2 {}_2\g(\vk,\vnhat,-1) =\:& \Bigg\{
- \frac14 
\left[ h_{-1}(\vk,\eta_0) \left(Q^*_1(x) e^{i x \mu}\right)\Big|_{x=0}
+ \left(1 - \frac23 C_1\rhocr H_0^{-2}\tilde a^{-2} \left\{\partial_{\tilde\eta}^2 + \tilde a\tilde H \partial_{\tilde\eta}\right\}\right)
h_{-1}(\vk,\tilde\eta) Q^*_1(\tilde x) e^{i \tilde x \mu }\right]
\vs
 & \quad + \int_0^{\chit} \frac{d\chi}{\chi}
\bigg[ \hat Q^*_2(x) + \frac{\chi}{\chit} \hat Q^*_3(x)
\bigg]
e^{i x \mu} h_{-1}(\vk,\eta_0-\chi)
\Bigg\}(1-\mu^2) e^{-i2\phi}.
\ea
Similarly, we can derive the corresponding expressions for
${}_{-2}\g$,
\ba
{}_{-2}\g(\vk,\vnhat,+1) =\:&
- \frac14 \left[ h_1(\vk,\eta_0) e^{i x\mu}\Big|_{x=0} + 
\left(1 - \frac23 C_1\rhocr H_0^{-2}\tilde a^{-2} \left\{\partial_{\tilde\eta}^2 + \tilde a\tilde H \partial_{\tilde\eta}\right\}\right)
h_1(\vk,\tilde\eta) e^{i \tilde x \mu}\right] (1 + \mu)^2 e^{i2\phi} \vs
 & + \int_0^{\chit} d\chi
\bigg[
\frac{\chit-\chi}{4}\frac{\chi}{\chit}
k^2 (1-\mu^2)^2
+ \left(1-2\frac{\chi}{\chit}\right) 
\frac{-ix}{2\chi} (1-\mu^2) (-\mu-1)
+ \frac1{2\chit}(1 + \mu)^2 
\bigg] e^{i2\phi} h_1(\vk,\eta_0-\chi) e^{i x \mu},
\ea
and correspondingly for $p=-1$,
by acting twice with the spin-raising operator for $m=\pm2$,
$(-\partial_\mu \mp 2/(1-\mu^2))$.  This immediately leads to
\ba 
\Del^2 {}_{-2}\g(\vk,\vnhat,+1) =\:& \Bigg\{
- \frac14 
\left[ h_1(\vk,\eta_0) \left(Q^*_1(x) e^{i x \mu}\right)\Big|_{x=0}
+ \left(1 - \frac23 C_1\rhocr H_0^{-2}\tilde a^{-2} \left\{\partial_{\tilde\eta}^2 + \tilde a\tilde H \partial_{\tilde\eta}\right\}\right)
h_1(\vk,\tilde\eta) Q^*_1(\tilde x) e^{i \tilde x \mu }\right]
\vs
 & \quad + \int_0^{\chit} \frac{d\chi}\chi
\bigg[ \hat Q^*_2(x) + \frac{\chi}{\chit} \hat Q^*_3(x) \biggl]
e^{i x \mu} h_1(\vk,\eta_0-\chi)
\Bigg\}(1-\mu^2) e^{i2\phi} 
\vs
%%%
\Del^2 {}_{-2}\g(\vk,\vnhat,-1) =\:& \Bigg\{
- \frac14 
\left[ h_{-1}(\vk,\eta_0) \left(\hat Q_1(x) e^{i x \mu}\right)\Big|_{x=0}
+ \left(1 - \frac23 C_1\rhocr H_0^{-2}\tilde a^{-2} \left\{\partial_{\tilde\eta}^2 + \tilde a\tilde H \partial_{\tilde\eta}\right\}\right)
h_{-1}(\vk,\tilde\eta) \hat Q_1(\tilde x) e^{i \tilde x \mu }\right]
\vs
 & \quad + \int_0^{\chit} \frac{d\chi}{\chi}
\bigg[ \hat Q_2(x) + \frac{\chi}{\chit} \hat Q_3(x)
\bigg]
e^{i x \mu} h_{-1}(\vk,\eta_0-\chi)
\Bigg\} (1-\mu^2) e^{-i2\phi} .
\ea
This is in the desired form of Eq.~(A17) in App.~A1 of \cite{stdruler},
with azimuthal harmonic index $r=\pm 2$.  Since the shear ${}_{\pm2}\g$
is a spin$\pm2$ quantity, and $h_{\pm 1}$ are two
independent polarization states with power spectra 
$P_{h_{\pm 1}}(k) = P_{T0}(k)/8$, we can apply Eq.~(A24) in \cite{stdruler}
with $s=2$, $r=2$, $N_P=2$, and $P_h = P_{T0}/8$:
\ba
C_\g^{XX}(l) =\:& \frac1{2\pi} 
\int k^2 dk\: P_{T0}(k) |F_l^{\g X}(k)|^2 \label{eq:Clshear} \\
%%%
F_l^{\g E}(k) \equiv\:& 
- \frac14 
\left[ T_T(k,\eta_0) \left(\Re \hat Q_1(x) \frac{j_l(x)}{x^2}\right)_{x=0}
+ \left(1 - \frac23 C_1\rhocr H_0^{-2}\tilde a^{-2} \left\{\partial_{\tilde\eta}^2 + \tilde a\tilde H \partial_{\tilde\eta}\right\}\right)
T_T(k,\tilde\eta)\Re \hat Q_1(\tilde x) \frac{j_l(\tilde x)}{\tilde x^2}\right] \vs
 & + \int_0^{\chit} \frac{d\chi}\chi
\bigg[ \Re \hat Q_2(x) + \frac{\chi}{\chit} \Re \hat Q_3(x)
\bigg]
\frac{j_l(x)}{x^2} T_T(k, \eta_0-\chi)
\vs
%%%
F_l^{\g B}(k) \equiv\:& 
- \frac14 
\left[ T_T(k,\eta_0) \left(\Im \hat Q_1(x) \frac{j_l(x)}{x^2}\right)_{x=0}
+ \left(1 - \frac23 C_1\rhocr H_0^{-2}\tilde a^{-2} \left\{\partial_{\tilde\eta}^2 + \tilde a\tilde H \partial_{\tilde\eta}\right\}\right)
T_T(k,\tilde\eta)\Im \hat Q_1(\tilde x) \frac{j_l(\tilde x)}{\tilde x^2}\right] \vs
 & + \int_0^{\chit} \frac{d\chi}\chi
\bigg[ \Im\hat Q_2(x) + \frac{\chi}{\chit} \Im\hat Q_3(x)
\bigg] \frac{j_l(x)}{x^2} T_T(k, \eta_0-\chi).
\nonumber
\ea
Again, $\tilde\eta = \eta_0-\chit$, $\tilde a = a(\tilde\eta)$, 
and $x=k\chi,\,\tilde x = k\chit$.  
This completes the derivation of the angular power spectrum of
$E$- and $B$-modes of the shear.  The operators $\hat Q_i$ when
applied to spherical Bessel functions can be simplified to yield
\ba
\Re \hat Q_1(x) \frac{j_l(x)}{x^2} =\:& -\frac1{x^2} \left[ 
(2x^2 - l^2 - 3l -2) j_l(x) + 2x j_{l+1}(x)
\right]
\vs
\Im \hat Q_1(x) \frac{j_l(x)}{x^2} =\:& -\frac1x \left[
2 (l+2) j_l(x) - 2x j_{l+1}(x) 
\right]
\vs
=\:& 2 \left[(l-1) \frac{j_l(x)}{x} - j_{l-1}(x) \right] \vs
\Re \hat Q_2(x) \frac{j_l(x)}{x^2} =\:& 
-\frac14 \left[(l^4 - 5l^2+4) \frac{j_l(x)}{x^2} + 2 (l^2+l-2) \frac{j_{l+1}(x)}{x}\right]
\vs
=\:& -\frac14 (l+2)(l-1)
\left[(l+1)(l-2) \frac{j_l(x)}{x^2} + 2 \frac{j_{l+1}(x)}{x}\right] \vs
\Im \hat Q_2(x) \frac{j_l(x)}{x^2} =\:& - \frac{(l-1)(l+2)}{2} \frac{j_l(x)}{x}
\vs
\Re \hat Q_3(x) \frac{j_l(x)}{x^2} =\:& \frac14 \frac{(l+2)!}{(l-2)!} \frac{j_l(x)}{x^2}
\vs
\Im \hat Q_3(x) \frac{j_l(x)}{x^2} =\:& 0.
\label{eq:Qibessel}
\ea
In the limit of $x \to 0$ for $l=2$, we have
\ba
\Re \hat Q_1(x) \frac{j_2(x)}{x^2} \stackrel{x\to 0}{=}\:& \frac45 \vs
\Re \hat Q_3(x) \frac{j_2(x)}{x^2} \stackrel{x\to 0}{=}\:& \frac25,
\ea
while all other operators vanish in this limit for $l = 2$, and all operators
vanish in this limit for $l>2$.  
With this, we can easily verify that modes with $k\to 0$ do not
contribute to the quadrupole, as desired.  
As we let $k\to 0$, and thus $x\to 0$, we trivially have $F_l^{\g B}(k)\to 0$, and
\ba
F_l^{\g E}(k) \stackrel{k\to 0}{=} 
- \frac14 2\left(\frac45\right)
 + \int_0^{\chit} \frac{d\chi}{\chi}
\frac{\chi}{\chit} \left(\frac25\right)
= -\frac25 + \frac25 = 0,
\ea
where we have used that $T_T(k\to 0,\eta) \to 1$ (of course, we only need
the fact that $T_T(k\to 0,\eta) \to $~const).

%%%%%%%%%%%%%%%%%%%%%%%%%%%%%%%%%%%%%%%%%%%%%%%%%%%%%%%%%%%%%%%%%%%%%%%%
\section{Connection to convergence and rotation}
\label{app:kom}

In this section, we cross-check \refeq{Clshear} with the
angular power spectra of coordinate convergence and rotation, through
the relations \refeq{CgCk}.  

% % % % % % % % % % % % % %% % % % % % % % % % % % %%% %  % %  % % % % %
\subsection{Angular power spectrum of coordinate convergence}
\label{app:kappa}

We begin with the general expression for the coordinate convergence
$\hat\k \equiv -1/2 \partial_{\perp i}\D x_\perp^i$ derived in 
\cite{stdruler}, restricted to synchronous-comoving gauge, and assuming
a tranverse-traceless metric perturbation $h_{ij}$:
\ba
\hat\k =\:& \frac34 (h_\parallel)_o 
+\frac12 \int_0^{\chit} d\chi\,\Bigg[
- \partial_\parallel h_\parallel
- \frac{3}{\chi} h_\parallel
+ (\chit-\chi)\frac{\chi}{\chit}  \nabla_\perp^2\left\{
 - \frac12 h_\parallel \right\}
\Bigg] \vs
%%%
=\:& \frac34 (h_\parallel)_o 
+\int_0^{\chit} d\chi\,\left[
- \frac12 \partial_\parallel h_\parallel
- \frac{3}{2\chi} h_\parallel \right]
- \frac14 \nabla_\Omega^2 \int_0^{\chit} d\chi \frac{\chit-\chi}{\chi\chit} h_\parallel.
\ea
This is equivalent to the expression used in \cite{paperI}, but in a form
more convenient for the comparison with the shear.  
Considering a single plane-wave perturbation oriented along the $z$-axis,
with $+1$ circular polarization, we have
\ba
\hat\k(\vnhat,\vk,+1) =\:& \frac34 (1-\mu^2) e^{i2\phi} e^{i x \mu}\Big|_{x=0} 
h_1(\vk,\eta_0)
+\int_0^{\chit} d\chi\,\left[
- \frac1{2\chi} x \partial_x  - \frac{3}{2\chi} \right]
(1-\mu^2) e^{i2\phi} e^{i x \mu} h_1(\vk,\eta) \vs
& - \frac14 \nabla_\Omega^2 \int_0^{\chit} d\chi \frac{\chit-\chi}{\chi\chit} 
(1-\mu^2) e^{i2\phi} e^{i x \mu} h_1(\vk,\eta),
\ea
where we have used \refeq{econtr}, and turned $i\mu k$ into $x/\chi \partial_x$, 
understanding that
the derivative only acts on $e^{ix\mu}$.  In order to derive
the multipole moments of $\hat\k$, we now use the relation 
(see App.~A in \cite{stdruler})
\ba
\int d\Omega\: Y^*_{lm} (1-\mu^2) e^{\pm i 2 \phi} e^{i x \mu} 
=\:& -\sqrt{4\pi (2l+1)} \sqrt{\frac{(l+2)!}{(l-2)!}}\, i^l \frac{j_l(x)}{x^2} \d_{m \pm 2},
\label{eq:Ylmint}
\ea
which yields
\ba
a^{\hat\k}_{lm}(\vk) 
=\:& -\sqrt{\frac{2l+1}{4\pi}\frac{(l+2)!}{(l-2)!}} (4\pi) i^l \d_{m2} \Bigg[
\frac34 h_1(\vk,\eta_0) \frac{j_l(x)}{x^2}\Big|_{x=0} \vs
& \hspace*{4.6cm} + \int d\chi\,h_1(\vk,\eta) \left\{
-\frac1{2\chi} x\partial_x - \frac3{2\chi} 
+ \frac14 l (l+1) \frac{\chit-\chi}{\chi\chit} \right\}
\frac{j_l(x)}{x^2} \Bigg].
\ea
We thus obtain for the angular power spectrum of $\hat\k$:
\ba
C_{\hat\k}(l) =\:& \frac1{2\pi} \int k^2 dk\, P_{T0}(k) |F_l^{\hat\k}(k)|^2 \vs
F_l^{\hat\k}(k) =\:& \sqrt{\frac{(l+2)!}{(l-2)!}} \Bigg[
-\frac34 T_T(k,\eta_0) \frac{j_l(x)}{x^2}\Big|_{x=0} 
- \int \frac{d\chi}\chi\,T_T(k,\eta_0-\chi) \left\{
-\frac12 x\partial_x - \frac32
+ \frac14 l (l+1) \left(1- \frac\chi{\chit}\right) \right\}
\frac{j_l(x)}{x^2} \Bigg] \vs
=\:& \sqrt{\frac{l(l+1)}{(l+2)(l-1)}} \Bigg[
-(l+2)(l-1) \frac34 T_T(k,\eta_0) \frac1{15} \d_{l2} \vs
& \hspace*{2.5cm} - (l+2)(l-1)\int \frac{d\chi}\chi\,T_T(k,\eta_0-\chi) \left\{
-\frac12 x\partial_x - \frac32
+ \frac14 l (l+1) \left(1- \frac\chi{\chit}\right) \right\}
\frac{j_l(x)}{x^2} \Bigg],
\ea
where for the observer term 
we have used $\lim_{x\to 0} j_l(x)/x^2 = 1/15$ for $l=2$, and $0$
for $l > 2$.  
We can now simplify the operator applied to the spherical Bessel function,
using the recurrence relation $j_l' = l/x\,j_l - j_{l+1}$,
yielding
\ba
\left\{-\frac12 x\partial_x - \frac32 + \frac14 l (l+1) \right\}
\frac{j_l(x)}{x^2} =\:& -\frac12\left\{[l+1-\frac12 l(l+1)] \frac{j_l}{x^2}
- \frac{j_{l+1}}{x} \right\} \vs
=\:& \frac14 (l-2)(l+1) \frac{j_l}{x^2} + \frac12 \frac{j_{l+1}}{x}.
\ea
We thus have
\ba
F_l^{\hat\k}(k) =\:& \sqrt{\frac{l(l+1)}{(l+2)(l-1)}} \Bigg[
-\frac15 T_T(k,\eta_0) \d_{l2} \vs
& \hspace*{2.5cm} - \int \frac{d\chi}\chi\,T_T(k,\eta_0-\chi) 
\bigg[\frac{(l+2)(l-1)}4\left\{
(l-2)(l+1) \frac{j_l}{x^2} + 2 \frac{j_{l+1}}{x} \right\}
- \frac14\frac\chi{\chit}\frac{(l+2)!}{(l-2)!} \frac{j_l(x)}{x^2} \bigg]
\Bigg].
\ea
For comparison, the corresponding filter function for the $E$-mode
of the shear is (without metric shear and IA contributions, as discussed
in \refsec{kom})
\ba
F_l^{\g E}(k) =\:& -\frac15 T_T(k,\eta_0) \d_{l2} -
\int_0^{\chit} \frac{d\chi}\chi
\bigg[ \frac{(l+2)(l-1)}4 \left\{ (l+1)(l-2) \frac{j_l(x)}{x^2} 
+ 2 \frac{j_{l+1}(x)}{x}\right\}
 - \frac{\chi}{\chit} \frac14 \frac{(l+2)!}{(l-2)!} \frac{j_l(x)}{x^2}
\bigg]
 T_T(k, \eta_0-\chi),
\ea
where we have used the $x\to 0$ limit for $\Re \hat Q_1 j_l/x^2$ for
the observer term.  Hence,
\be
F_l^{\g E}(k) = \sqrt{\frac{(l+2)(l-1)}{l(l+1)}} F_l^{\hat\k}(k).
\ee
Our result for the shear (without metric shear and intrinsic alignment) 
thus recovers the full-sky relation between
shear $E$-(gradient-)modes and coordinate convergence \refeq{CgCk}.  

% % % % % % % % % % % % % %% % % % % % % % % % % % %%% %  % %  % % % % %
\subsection{Angular power spectrum of rotation}
\label{app:rot}

Using the definition of the rotation \refeq{kom}, and the expression for
$\D x_\perp^i$ [Eq.~(36) in \cite{stdruler}] restricted to synchronous
gauge, we obtain
\ba
\omega =\:& \frac12 \int_0^{\chit} d\chi \Big[
\eps_{ijk} \nhat^i (\partial_\perp^j h^k_{\;m}) \nhat^m \Big] \vs
=\:& \frac12 \int_0^{\chit} d\chi \Big[
\eps_{ijk}   h^{k\  ,j}_{\;m} 
\Big] \nhat^i \nhat^m.
\label{eq:om2}
\ea
Note that
$\eps_{ijk} \partial_\perp^j \nhat^k = \eps_{ijk} \partial^j \nhat^k = 0$,
and $\nhat^j \partial_{\perp\,j} f(\vx,\eta) = 0$,
and that pulling the derivatives inside the integrand yields a
factor of $\chi/\chit$.  
This result agrees with Eq.~(4) in \cite{DodelsonEtal} 
(this $\omega$ is also equivalent to $\nabla_\theta^2 \Omega$
as defined above Eq.~(19) of \cite{BookEtal}).  
We now calculate the
angular power spectrum of $\omega$.  Assuming as above
a single plane wave with $\vk = k \hat{\v{z}}$, we have
\ba
\eps_{ijk}   h^{k\  ,j}_{\;m} \nhat^i \nhat^m =\:&
- ik \left[ - 2\nhat^1 \nhat^2 h_+ + \left( (\nhat^1)^2 - (\nhat^2)^2\right)
h_\times \right] e^{i\vk\cdot\vx}
\vs
=\:& -ik (1-\mu^2) \left[ - \sin 2\phi h_+ 
+ \cos 2\phi h_\times \right] e^{i k\chi \mu}\vs
=\:& k (1-\mu^2) \left[ h_1 e^{2i\phi} - h_{-1} e^{-2i\phi} \right] e^{i k\chi \mu},
\ea
since $h_{-1} = h_1^*$.  Thus,
\be
\omega(\vnhat,\vk) = \frac12\int_0^{\chit} d\chi \: k
\left[ h_1 e^{2i\phi} - h_{-1} e^{-2i\phi} \right] (1-\mu^2) e^{i k\chi \mu}.
\ee
Note the relative minus sign between the two polarization states
which shows that $\omega$ is parity-odd.  In analogy with the derivation
for $\hat\k$ (see also App.~A1 in \cite{stdruler})
we use \refeq{Ylmint} to obtain the angular power spectrum
of the rotation (it only contains ``$B$-modes''):
\ba
C_\omega^{BB}(l) =\:& \frac1{2\pi} 
\int k^2 dk\: P_{T0}(k) |F_l^{\omega}(k)|^2 \label{eq:Comega} \\
F_l^\omega(k) \equiv\:& 
- \frac12 \sqrt{\frac{(l+2)!}{(l-2)!}}\int \frac{d\chi}{\chi}\: x\, T_T(k,\eta_0-\chi) \frac{j_l(x)}{x^2}\Big|_{x=k\chi}
\label{eq:Flom}.
\ea
In \citet{DodelsonEtal}, an additional ``metric shear'' term was
added to $F_l^\omega(k)$:
\ba
F_l^\omega \to\:& F_l^\omega + F_l^{\omega\,\rm MS} \vs
F_l^{\omega\,\rm MS}(k) =\:& -\frac12 \frac1{(l+2)(l-1)} \sqrt{\frac{(l+2)!}{(l-2)!}}
\left[(l-1) \frac{j_l(\tilde x)}{\tilde x} - j_{l-1}(\tilde x)\right] T_T(k,\tilde\eta).
\ea
Note that in our convention, there is an overall minus sign for both 
$F_l^\omega$ and $F_l^{\omega,\rm MS}$.  
Next, consider the $B$-mode power spectrum of the shear without observer, 
IA and FNC terms, \refeq{Clshear} with
\ba
F_l^{\g B}(k) =\:& -\frac12 (l+2)(l-1) \int_0^{\chit} \frac{d\chi}{\chi}
x\,T_T(k, \eta_0-\chi) \frac{j_l(x)}{x^2} \vs
=\:& \sqrt{\frac{(l+2)(l-1)}{l(l+1)}} F_l^{\omega}(k),
\ea
where we have used \refeq{Qibessel} and \refeq{Flom}.  
We thus recover the relation \refeq{CgCk} between the $B$-modes of the shear
and the rotation (without ``metric shear'').  
The contribution of the FNC term to the $B$-modes of the shear is
given by
\ba
F_l^{\g B\,\rm FNC}(k) =\:& -\frac14 T_T(k,\tilde\eta) \Im \hat Q_1(\tilde x)
\frac{j_l(\tilde x)}{\tilde x^2} = -\frac12 T_T(k,\tilde\eta)
\left[(l-1) \frac{j_l(\tilde x)}{\tilde x} - j_{l-1}(\tilde x)\right] \vs
=\:& \sqrt{\frac{(l+2)(l-1)}{l(l+1)}} F_l^{\omega\,\rm MS}(k),
\ea
showing that the ``metric shear'' contribution to $\omega$ 
derived in \cite{DodelsonEtal} agrees with the contribution of the FNC 
term to the shear $B$-modes.

%%%%%%%%%%%%%%%%%%%%%%%%%%%%%%%%%%%%%%%%%%%%%%%%%%%%%%%%%%%%%%%%%%%%%%%%%%%
\end{widetext}

%%%%%%%%%%%%%%%%%%%%%%%%%%%%%%%%%%%%%%%%%%%%%%%%%%%%%%%%%%%%%%%%%%%%%%%%%%%
%\bibliographystyle{arxiv_physrev}
\bibliography{GW}

\end{document}